\documentclass[twocolumn]{aastex631} % twocolumn

\usepackage{amsmath}
\usepackage{nicefrac}
\newcommand{\hst}{\textit{HST}}
\newcommand{\tess}{\textit{TESS}}

\newcommand{\rearth}{$R_\oplus$}
\newcommand{\mearth}{$M_\oplus$}

\newcommand{\rsun}{$R_\odot$}
\newcommand{\msun}{$M_\odot$}

\newcommand{\water}{$\mathrm{H}_2\mathrm{O}$}
\newcommand{\methane}{$\mathrm{CH}_4$}
\newcommand{\dioxide}{$\mathrm{CO}_2$}
\newcommand{\ammonia}{$\mathrm{NH}_3$}

\shorttitle{WFC3 Spectroscopy of TOI-674 b}
\shortauthors{Brande et al.}
\graphicspath{{./}{figures/}}
\begin{document}

%\title{Water in the Oasis: HST WFC3 NIR Spectroscopy of TOI-674 b}
\title{A Mirage or an Oasis? Water Vapor in the Atmosphere of the Warm Neptune TOI-674 b}

\author[0000-0002-2072-6541]{Jonathan Brande}
\affiliation{Department of Physics and Astronomy, University of Kansas, 1082 Malott, 1251 Wescoe Hall Dr., Lawrence, KS 66045, USA}
\correspondingauthor{Jonathan Brande}
\email{jbrande@ku.edu}

\author{Ian J. M. Crossfield}
\affiliation{Department of Physics and Astronomy, University of Kansas, 1082 Malott, 1251 Wescoe Hall Dr., Lawrence, KS 66045, USA}

\author[0000-0003-0514-1147]{Laura Kreidberg}
\affiliation{Max Planck Institute for Astronomy, K\"{o}nigstuhl 17, 69117 Heidelberg, Germany}

\author[0000-0002-9584-6476]{Antonija Oklopčić}
\affiliation{Anton Pannekoek Institute of Astronomy, University of Amsterdam, Science Park 904, 1098 XH Amsterdam, Netherlands}

\author[0000-0001-7047-8681]{Alex S. Polanski}
\affiliation{Department of Physics and Astronomy, University of Kansas, 1082 Malott, 1251 Wescoe Hall Dr., Lawrence, KS 66045, USA}

\author{Travis Barman         }
\affiliation{Lunar and Planetary Laboratory, University of Arizona, Tucson, AZ 85721 USA}

\author{Bj\"{o}rn Benneke        }
\affiliation{Departement de Physique, and Institute for Research on Exoplanets, Universite de Montreal, Montreal, Canada}

\author[0000-0002-8035-4778]{Jessie L. Christiansen}
\affiliation{Caltech/IPAC-NASA Exoplanet Science Institute, Pasadena, CA 91125, USA}

\author{Diana Dragomir        }
\affiliation{Department of Physics and Astronomy, University of New Mexico, Albuquerque, NM, USA}

\author[0000-0002-9328-5652]{Daniel Foreman-Mackey}
\affiliation{Center for Computational Astrophysics, Flatiron Institute, New York, NY 10010, USA}

\author[0000-0002-9843-4354]{Jonathan J. Fortney      }
\affiliation{Department of Astronomy and Astrophysics, University of California, Santa Cruz, CA 95064, USA}

\author[0000-0002-8963-8056]{Thomas P.\ Greene            }
\affiliation{NASA Ames Research Center Space Science and Astrobiology Division M.S. 245-6 Moffett Field, CA 94035, USA}

\author[0000-0001-8638-0320]{Andrew W. Howard         }
\affiliation{Cahill Center for Astronomy \& Astrophysics, California Institute of Technology, Pasadena, CA 91125, USA}

\author[0000-0002-5375-4725]{Heather A. Knutson       }
\affiliation{Division of Geological and Planetary Sciences, California Institute of Technology, Pasadena, CA 91125, USA}

\author[0000-0003-3667-8633]{Joshua D. Lothringer     }
\affiliation{Physics Department, Utah Valley University, 800 West University Parkway, Orem, UT 85058-5999, USA}

\author{Thomas Mikal-Evans    }
\affiliation{Max Planck Institute for Astronomy, K\"{o}nigstuhl 17, 69117 Heidelberg, Germany}

\author[0000-0002-4404-0456]{Caroline V. Morley       }
\affiliation{Department of Astronomy, University of Texas at Austin, Austin, TX, USA}

% Other collaborators?




\begin{abstract}
We report observations of the recently discovered warm Neptune TOI-674 b (5.25 \rearth{}, 23.6 \mearth{}) with the Hubble Space Telescope's Wide Field Camera 3 instrument. TOI-674 b is in the Neptune desert, an observed paucity of Neptune-size exoplanets at short orbital periods. Planets in the desert are thought to have complex evolutionary histories due to photoevaporative mass loss or orbital migration, making identifying the constituents of their atmospheres critical to understanding their origins. We obtained near-infrared transmission spectroscopy of the planet's atmosphere with the G141 grism. After extracting, detrending, and fitting the spectral lightcurves to measure the planet's transmission spectrum, we used the petitRADTRANS atmospheric spectral synthesis code to perform retrievals on the planet's atmosphere to identify which absorbers are present. These results show moderate evidence for increased absorption at 1.4 $\mu$m due to water vapor at 2.9$\sigma$ (Bayes factor = 15.8), as well as weak evidence for the presence of clouds at 2.2$\sigma$ (Bayes factor = 4.0). TOI-674 b is a strong candidate for further study to refine the water abundance, which is poorly constrained by our data. We also incorporated new TESS short-cadence optical photometry, as well as Spitzer/IRAC data, and re-fit the transit parameters for the planet. We find the planet to have the following transit parameters: $R_p/R_* = 0.1135\pm0.0006$, $T_0 = 2458544.523792\pm0.000452$ BJD, and $P = 1.977198\pm0.00007$ d. These measurements refine the planet radius estimate and improve the orbital ephemerides for future transit spectroscopy observations of this highly intriguing warm Neptune.

\end{abstract}



\section{Introduction} \label{sec:intro}

The last two and a half decades of exoplanet science have revealed a wealth of information on planetary system architectures. The first discovered exoplanet around a main-sequence star star, 51 Pegasi b \citep{mayor1995} is a Hot Jupiter, one of a class of planets that challenged our ideas on the formation and evolution of planetary systems. As the field has progressed, these astonishing outliers have proven to be representative of larger planetary populations in systems often unlike our own. 

In addition to these populations, several gaps in the distribution of short-period exoplanets have also been noted, namely the radius valley \citep{fulton2017} and the Neptune desert \citep{mazeh2016}. For the radius valley, atmospheric mass loss due to host star irradiation is the main theory for the observed lack of 1.5--2 \rearth{} planets at these short orbital periods \citep{owen2017}. Other explanations due to formation mechanisms and core-powered mass loss \citep{ginzburg2018} have also been put forth, as well as a primordial radius gap due to late gas accretion in gas-poor nebulae \citep{lee2021}. The Neptune desert is a similar lack of planets at even shorter orbital periods ($P \leq 2-4$~d) but for approximately Neptune-to-Jupiter mass planets \citep{mazeh2016}. The lower-mass section of the gap may be appropriately explained by irradiative atmospheric stripping, but the dearth of Jupiter-mass planets in this narrow period range may be better explained by planetary migration and in-situ formation \citep{owen2018, bailey2018}.

The Neptune desert is especially relevant, given the uncertainties in our own solar system about the formation of Uranus and Neptune, either through core accretion \citep{frelikh2017} or disk instability \citep{boss2003}. We presume migration processes were important in their early histories as they would have been for Jupiter and Saturn, and by proxy also the observed exoplanetary populations. Transit surveys are not generally sensitive to these cold giant planets, and efforts to measure the true frequency of Solar system analogs rely on long-baseline radial velocity surveys \citep{wittenmyer2020}. However, these surveys still do not have the time baselines or RV precision to detect Uranus and Neptune-like planets, which would require $\sim 1$m/s RV precision, aided by $\mu$as astrometry from \textit{Gaia} \citep{wittenmyer2020}. %However, compared to the current observed sample of exoplanets, we find very few exoplanets with similar masses and orbital separations as Uranus and Neptune have in our own solar system, although this is certainly incomplete due to a lack of observational sensitivity for low-mass, widely separated planets. Most known planets in this mass range (10--40 \mearth{}) are relatively evenly distributed at intermediate separations, but populations at short orbital separations and wide orbital separations (including Uranus and Neptune) seem similarly sparse. %Observational capabilities certainly bias the sample, but  \textbf{DO WE HAVE ANYTHING ON M-DWARF WIDE PLANET COMPLETENESS?}

\begin{figure*}[htbp]%[htb]
    \centering
    \includegraphics[width=0.6\textwidth]{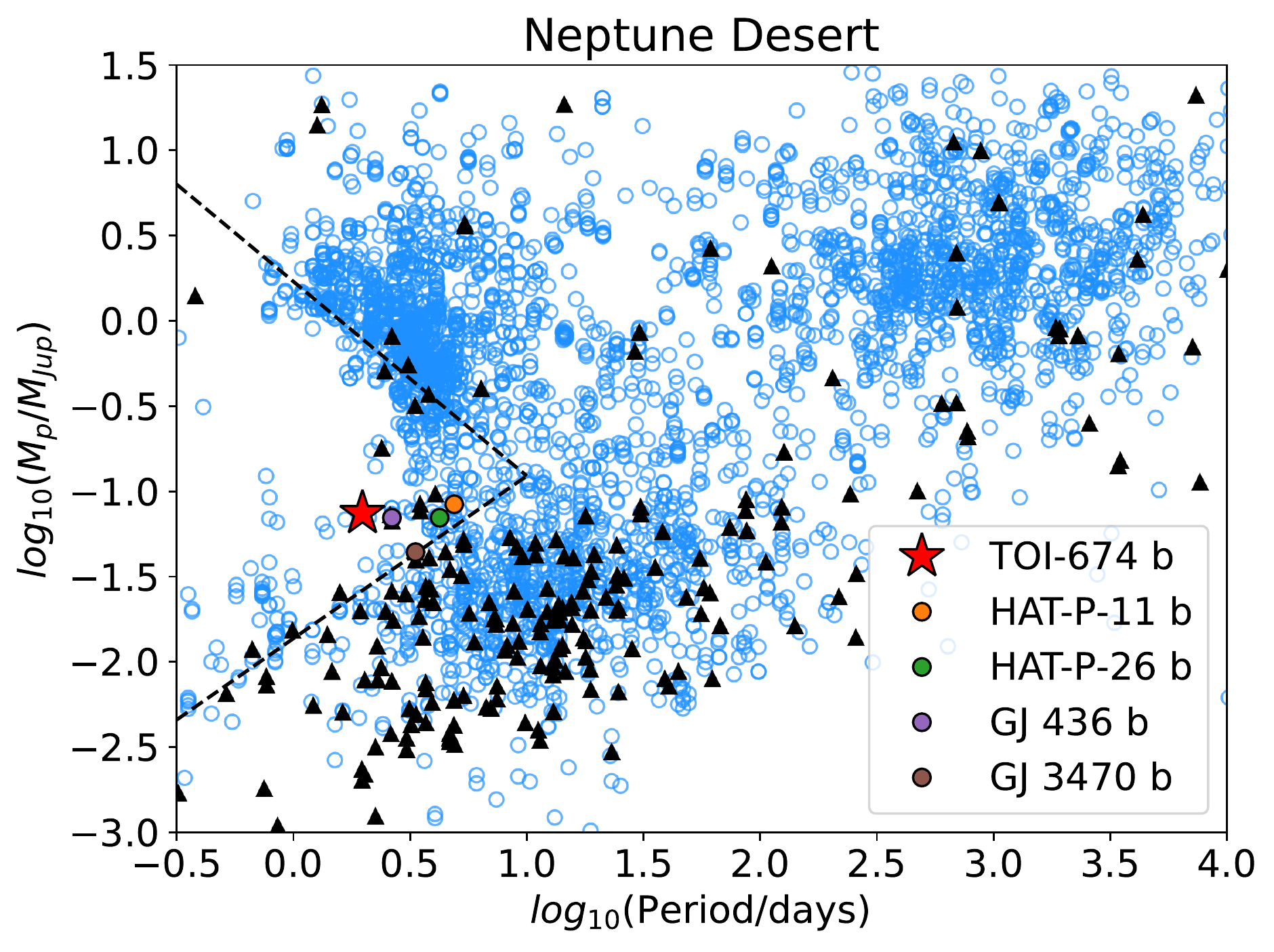}
    \vspace{-1em}
    \caption{Planet mass vs period for all planets with known masses and periods. The dashed black lines show the boundaries of the Neptune desert from \citep{mazeh2016}, the black triangles show M-star planets, the blue circles show all stellar hosts, and the red star shows TOI-674 b. We also show several other similar planets to TOI-674 b on the plot. The total planet sample has been significantly expanded since \citet{mazeh2016}, and the original sharply pointed boundary may in fact have a much more gradual limit near $10^{0.5}$ days. Planet mass and period data taken from the Exoplanet Archive \citep{exoarchive}}
%    \vspace{-1em}
    \label{fig:desert}
\end{figure*}

As can be seen in Fig.~\ref{fig:desert}, fewer total planets with measured masses are known to orbit M-dwarfs than other stellar types, making it difficult to say with certainty whether the Neptune desert exists around the coolest stars. As the upper boundary of the Neptune desert is characterized by planets with masses $\lesssim M_{Jup}$, the upper bound for the M-dwarf Neptune desert is unclear given the general lack of massive planets around M-dwarfs. However, the lower boundary of the desert appears to hold for the M-dwarf planet population.%, putting TOI-674 b fairly deep into the desert for the M-dwarf planet sample.

%As orbital period here is being used as a rough proxy for incident stellar flux, it is tempting to convert the boundaries of the Neptune desert to insolation to more directly compare the non-M-star and M-star planet samples. However, since there is no ``typical" stellar flux (although one might choose $1\ L_\odot$ or something similar as a reference value), comparing very intrinsically faint stars like M-stars and much more luminous stars like FGK-stars would be hard to do in a consistent and informative way. Given this and the tentative agreement of the M-star planet sample with the traditional desert boundaries, we use in this paper the traditional boundary definitions.

%% INSERT MASS/SMA PLOT OF ALL EXOPLANETS WITH NEPTUNE DESERT INDICATED AND SOLAR SYSTEM PLANETS PLOTTED

It is especially tempting to want to characterize the few large planets known to exist in the desert. Several high profile planet discoveries have been made in the Neptune desert \citep{bakos2010, borucki2010, hartman2011, bonomo2014, bakos2015, crossfield2016, eigmuller2017, barragan2018, west2019, jenkins2020}, and these planets may be exceptional in several ways. Young Neptune desert planets may be undergoing atmospheric mass loss, or may be in the process of migrating into the desert. Older Neptune desert planets may have already lost parts of their atmospheres, or finished their migrations. However, as this is only a relatively recently identified population, only a few have been characterized by atmospheric transmission spectroscopy (GJ 436 b: \citet{knutson2014b}, HAT-P-26 b: \citet{wakeford2017}, GJ 3470 b: \citet{benneke2019}, HAT-P-11 b: \citet{chachan2019}). These atmospheres range from featureless (GJ 436 b) to strongly featured (HAT-P-26 b), and also have varying metallicity, with both low metallicities (HAT-P-11 b, HAT-P-26 b and GJ 3470 b) and ambiguous metallicities (GJ 436 b), where both high and low metallicities could produce the observed atmosphere. %As atmospheric metallicity can be a proxy for planetary formation history, more data is needed on a greater diversity of targets to measure these or identify which targets can be further explored in the future.} 
Here we present Hubble Space Telescope (\hst{}) Wide Field Camera 3 (WFC3) infrared (IR) spectroscopic observations of the recently discovered warm Neptune TOI-674 b.

\subsection{TOI-674 b}
The TESS observatory recently discovered TOI-674 b, a warm Neptune (5.25 \rearth, 23.6 \mearth) orbiting a nearby M2 dwarf (TIC 158588995, V=14.2 mag, J=10.4 mag, RA 10$^\text{h}$58$^\text{m}$20.98$^\text{s}$ DEC -36$^{\circ}$51$'$29.13$''$ (J2000), 46.16 pc, 0.420 \rsun, 0.420 \msun) with a period of 1.977143 days \citep{murgas2021}. With these parameters, TOI-674 b is in the Neptune desert (see Fig. \ref{fig:desert}), experiencing 38 times as much stellar radiation as the Earth does \citep{murgas2021}.%, insert figure from HST proposal)}, an observed paucity of Neptune-to-Jupiter sized planets at short-to-intermediate orbits and high insolations. Two theories likely explain the presence of this desert: stellar wind atmospheric stripping for the lower-mass half of the desert, and migration mechanisms for the upper half. \textbf{Ask Eric Lopez about mass loss timescales.}

TOI-674 b also provides a good target for atmospheric transmission spectroscopy, which attracted our attention during the first year of the TESS mission. Given the small size of the host star and relatively large radius of the planet, TOI-674 b has a high transmission spectroscopy metric of 222 (see \citet{kempton2018} for a definition of this quantity). Compared to other similar planets in the desert \citep[see e.g.\ Fig.\ 10 in ][]{murgas2021}, these factors make it one of the best planets of its class for transmission spectroscopy.

In Section \ref{sec:data}, we describe our data and analyses, including details of our transit and systematics models for the \hst{} WFC3 data, as well as the \tess{} and \textit{Spitzer} data. In Section \ref{sec:retrieval}, we present the details and results of the atmospheric retrieval framework used here, and in Section \ref{sec:discussion}, we discuss the implications of these results, including future observations.

\section{Data, Data Reduction, and Analysis} \label{sec:data}
\subsection{Observations}
We observed three transits of TOI-674 b on 10, 12, and 26 July 2020 with the \textit{Hubble Space Telescope's} (\hst{}) Wide Field Camera 3 instrument, as part of the large \hst{} General Observer Program 15333 (Co-PIs: Crossfield and Kreidberg). Each transit visit consisted of four orbits, and each orbit started with one direct image in the F130N filter, and then continued with spectroscopic imaging with the G141 grism ($1.04 \mu$m -- $1.77 \mu$m). 55 exposures were taken in Visit 1, 60 exposures were taken in Visit 2, and 52 exposures were taken in Visit 3. All spectral exposures were taken with round-trip spatial scanning \citep{mccullough2012, deming2013}, and each scan direction had an exposure time of 134 seconds. The scan rate was 0.043 arcsec/s, and the scan length was 6.08 arcsec. All \hst{} data analyzed in this paper can be found in MAST at: \dataset[10.17909/tvy9-7h80]{http://dx.doi.org/10.17909/tvy9-7h80}. %\textbf{TALK ABOUT SCAN RATE, OBSERVING EFFICIENCY}. 

As TOI-674 b was discovered by the Transiting Exoplanet Survey Satellite (\tess{}) \citep{ricker2015}, we also have access to planet transit data in the \tess{} bandpass. The discovery paper was based on 22 transits observed in \tess{} Sectors 9 and 10 \citep{murgas2021}, and we use new photometry of the planet consisting of 11 transits in \tess{} Sector 36. Finally, we also observed a single transit of TOI-674 b with the \textit{Spitzer} Space Telescope's IRAC instrument in the $4.5 \mu$m channel (also incorporated into \citep{murgas2021}) Further discussion of these observations is presented in Section \ref{sec:spitzer}. We incorporate both the \tess{} and \textit{Spitzer} transit depths into our eventual atmospheric retrievals (see Sec. \ref{sec:retrieval}).

\subsection{Data Reduction and Analysis}
We used the \texttt{Iraclis} pipeline \citep{tsiaras2016a, tsiaras2016b, tsiaras2018} to reduce the raw spatially scanned \hst{} data and extract spectral lightcurves. \texttt{Iraclis} performs a standard set of \hst{} WFC3 image reduction steps (e.g. calibrating, flat-fielding, bad pixel/cosmic ray correction, etc.) and then extracts the spectrum from the reduced images.A full description of the \texttt{Iraclis} reduction steps are given in \citet{tsiaras2016a, tsiaras2016b, tsiaras2018}. %The \texttt{Iraclis} data reduction steps are as follows: zero-read subtraction, reference pixel correction, nonlinearity correction, dark current subtraction, gain conversion, sky background subtraction, calibration, flat-field correction, and bad pixel/cosmic ray correction. 
\texttt{Iraclis} ingests \hst{} flat-fielded direct images of the target star to locate the target on the detector, and then extracts the spatially scanned spectrum from the raw spectral data files. After conducting the reduction and extraction, \texttt{Iraclis} returns the reduced images and extracted spectra, along with some diagnostic information.
%Given the direct images in the FLT format, and the spectroscopic images in the RAW format, \texttt{Iraclis} will reduce the data and extract the spectrally-binned lightcurves as well as return some diagnostics in a convenient, Python-ingestable pickle file. 
Input parameter files allow a user to modify various aspects of the reduction, extraction, and fitting process. In order to determine the optimal extraction aperture to minimize scatter in the spectrophotometric lightcurves, we ran \texttt{Iraclis} with varying extraction apertures from 0 pixels above and below the spectrum to 20 pixels above and below the spectrum in 5 pixel increments. We then fit the broadband lightcurves for each extraction aperture and logged the RMS error for each. The 10 pixel aperture yielded the lowest RMS error, and we used this aperture for our extraction. We extracted 18 spectral bins ranging from $1.1108~\mu$m to $1.6042~\mu$m, such that each bin contains approximately equal stellar flux. The extracted lightcurves were then used as inputs for our transit model and systematics fitting process.

\subsubsection{Transit and Systematics Models}
HST/WFC3 lightcurves offer precise transit measurements but are known to be subject to significant systematic effects. In order to detrend the transit lightcurves we modified the \texttt{model-ramp} method from \citet{kreidberg2014} to fit our data. Our modification of the \texttt{model-ramp} method fits the systematics and the transit parameters simultaneously as follows: 

\begin{equation}
\begin{aligned}
M_{\lambda v}(t) = & F_{\lambda v}[M_{0,\lambda}(t)(1+V_{\lambda v} \mathbf{t_v})(1 - R_{\lambda v}e^{-\mathbf{t_v} / \tau_\lambda}) \\
& + (S_{\lambda v o} \cos \frac{\pi \mathbf{t_b}}{\tau_c})] \\
\end{aligned}
\end{equation}

%\textbf{SIT DOWN WITH CODE AND REWRITE}

$M_{\lambda v}(t)$ is the full model to the observed data, $F_{\lambda v}$ is the out-of-transit mean flux, $M_{0,\lambda}(t)$ is the bare normalized transit lightcurve, $V_{\lambda v}$ is a visit-long slope, $R_{\lambda v o}$ is the amplitude of the ramp systematic, $\tau_\lambda$ is the ramp systematic timescale, $\mathbf{t_v}$ is a vector of the times elapsed since the first exposure in the current visit, $S_{\lambda v o}$ the amplitude of the scan-direction sinusoid, $\mathbf{t_b}$ a vector of the times elapsed since the first exposure in the current orbit, and $\tau_c$ the average duration between the start of each exposure. $\lambda, v, o$ are subscripts denoting spectral bin central wavelength, \hst{} visit, and orbit number. We also include an extra error term $\sigma_F$, added in quadrature to the per-integration flux uncertainty, which we find aids in sampling.

Following previous analyses \citep{deming2013, wakeford2016, zhou2017, alderson2022}, we discard the initial orbit in each visit due to the strong effect of the ramp systematic in that orbit, and we also found that the initial spectral exposure in each orbit was also strongly affected by the ramp systematic and discarded it as well. Using the \texttt{exoplanet} toolkit \citep{dfm2021}, we fit the white-light transit lightcurves for each transit, fitting the exponential orbit-level ramp and visit-long slope systematics models as described in \citet{kreidberg2014}, and correcting for the round-trip scan effect with a sinusoidal model. We describe our method for correcting the spatial scan systematic in more detail in Appendix \ref{app:scan}, as well as benchmark it against legacy methods. The transit lightcurve and systematics parameters are normalized such that the out of transit flux is 1, and then the entire model is multiplied by the mean out of transit flux observed in the \hst{} data. 

Our \hst{} transit model incorporated the published star and planet parameters from \citet{murgas2021}, except where those parameters were refined by our new fit to the \tess{} data incorporating Sector 36. The stellar and planetary parameter priors are shown in Table~\ref{tab:mcpriors}. Limb darkening coefficients for the broadband transit and spectral bins were pre-calculated using the Limb Darkening Calculator in the Exoplanet Characterization Toolkit (ExoCTK) \citep{bourque2021}, using the published stellar parameters from \citet{murgas2021}, and the Kurucz ATLAS9 stellar models. %\textbf{WHAT'S THE SPECIFIC VERSION USED IN EXOCTK?}. 
All parameter estimation was conducted with the \texttt{exoplanet} toolkit \citep{dfm2021}, built on top of PyMC3 \citep{pymc2016} for posterior sampling. \texttt{exoplanet} uses gradient-based inference methods to improve sampling performance compared to ensemble samplers or nested sampling, which are more commonly used in astronomy. Here we use \texttt{exoplanet}'s No U-Turn Sampler \citep{hoffman2011} implementation. \texttt{exoplanet} also allows for the simulation of transit lightcurves using \texttt{starry} \citep{luger2019}. For each of the \hst{} transit fits (the white lightcurve and each spectral lightcurve, for each visit), we ran 4 chains for 2000 tuning steps (analogous to traditional MCMC burn-in, but the sampler adjusts the step sizes to better fit the gradient of the log-probability of instead of hopefully exiting a bad starting point) and then drew 4000 samples from which to construct our posterior distributions.

For the first transit observed, an example white-light transit fit can be seen in Figure \ref{fig:white_lc}. The white-light fits for transits 2 and 3 can be found in Appendix \ref{app:a}.

\begin{figure*}[ht]
    \centering
    \includegraphics[width=0.6\textwidth]{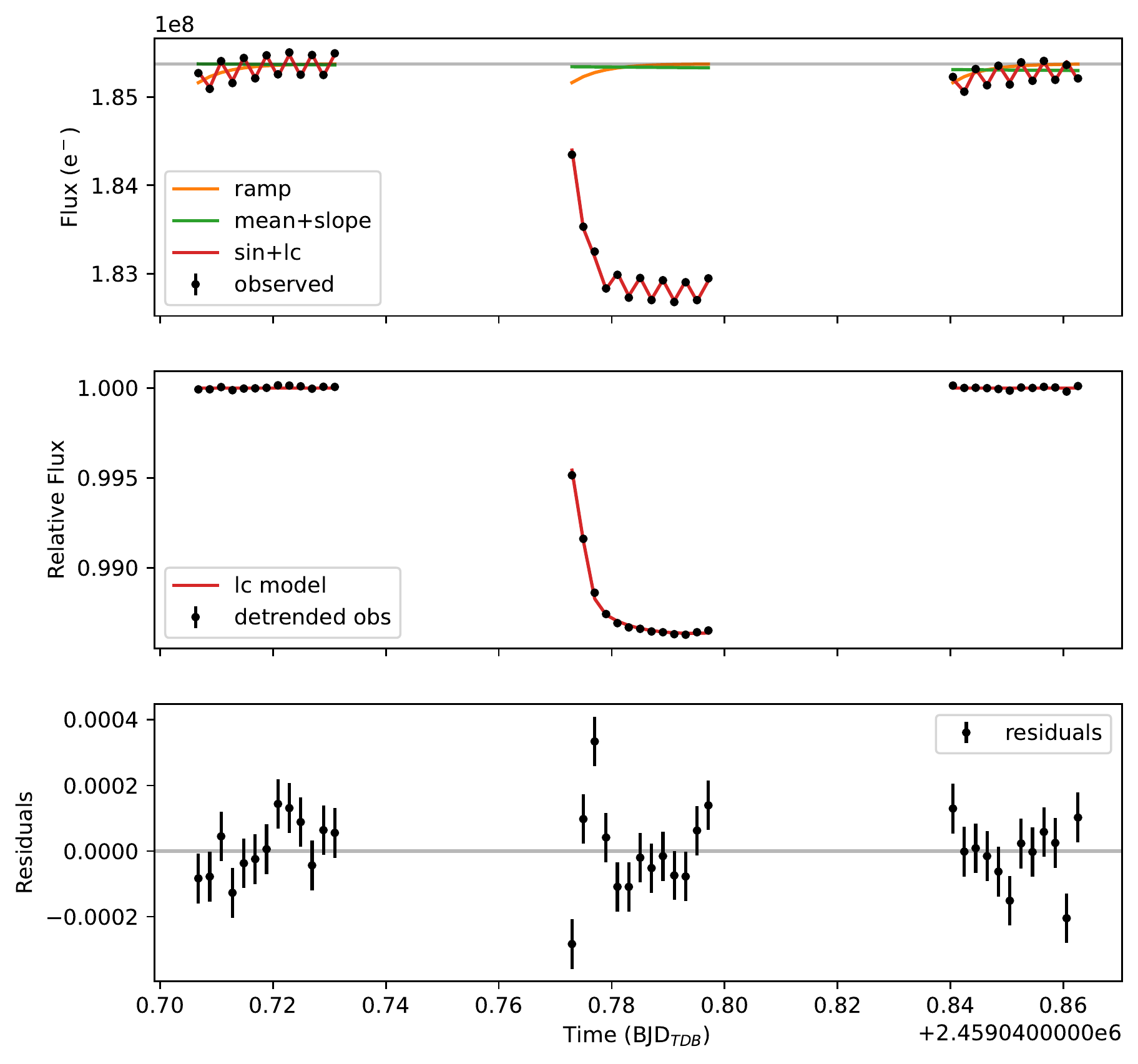}
    \vspace{-1em}
    \caption{The broadband data for the first transit of TOI-674 b. \textit{Top:} The raw transit data, with the systematics and transit model. \textit{Middle:} The detrended transit data and lightcurve model. \textit{Bottom:} The white-lightcurve residuals.}
%    \vspace{-1em}
    \label{fig:white_lc}
\end{figure*}

\begin{deluxetable}{lcl}[htb]
\tablecolumns{3}
\tablecaption{LC and Systematics Fitting Priors \label{tab:mcpriors}}
\tablehead{
\colhead{Parameter} & \hspace{1em} &\colhead{Prior} \\
}
\startdata
$T_0$ (BJD) &\hspace{10em} & $\mathcal{N}(2458641.405, 0.0104)$ \\
$r_p / r_*$ & & Lognormal(ln(0.1135), ln(0.001))\\
*$P$ (d) & & 1.977198 \\
*$e$ & & 0.0 \\
*$\omega$ & & 0.0 \\
$F_{\lambda v}$ & & Lognormal($\mu_{oot}$, ln($\sigma_F$)) \\
$V_{\lambda v}$      &       & $\mathcal{N}(0, 0.001)$ \\ 
$R_{\lambda v o}$   &       & $\mathcal{N}(0.001, 0.0001)$ \\
$\tau_\lambda$   &       & LogNormal($\frac{\tau_0}{2}$, 1.0) \\
$S_{\lambda v o}$  &       & $\mathcal{U}(0, 1)$ \\
$\sigma_F$ &        & InverseGamma(1800, 100)
\enddata
\tablecomments{Transit parameters used for the fitting priors were taken from our re-analysis of the full \tess{} transit dataset of TOI-674 b. The prior on $T_0$ was chosen by eye to correspond with the first \hst{} transit. Rows marked with * denote fixed values in the transit fit. $\mu_{oot}$ is the mean out-of-transit flux, $\sigma_F$ is $\frac{max(F_{oot}) - min(F_{oot})}{4}$, and $\tau_0$ is the duration of the first visit.}
\end{deluxetable}

\begin{figure}[htb]
    \centering
    \includegraphics[width=0.45\textwidth]{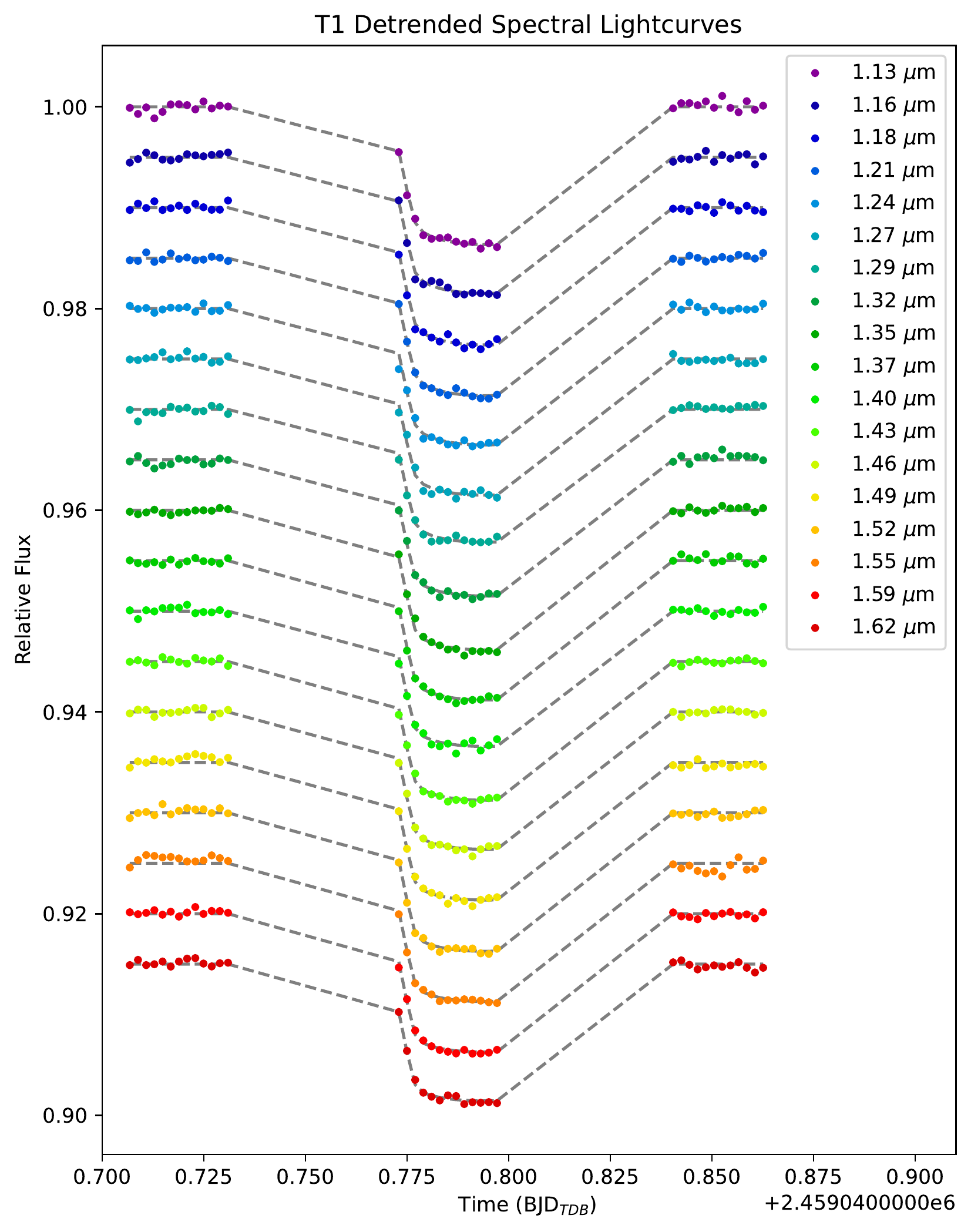}
    \vspace{-1em}
    \caption{Detrended spectral lightcurves and the transit models for the first transit of TOI-674 b.}
%    \vspace{-1em}
    \label{fig:t1_detrend}
\end{figure}

\begin{figure*}[htb]
    \centering
    \includegraphics[width=0.99\textwidth]{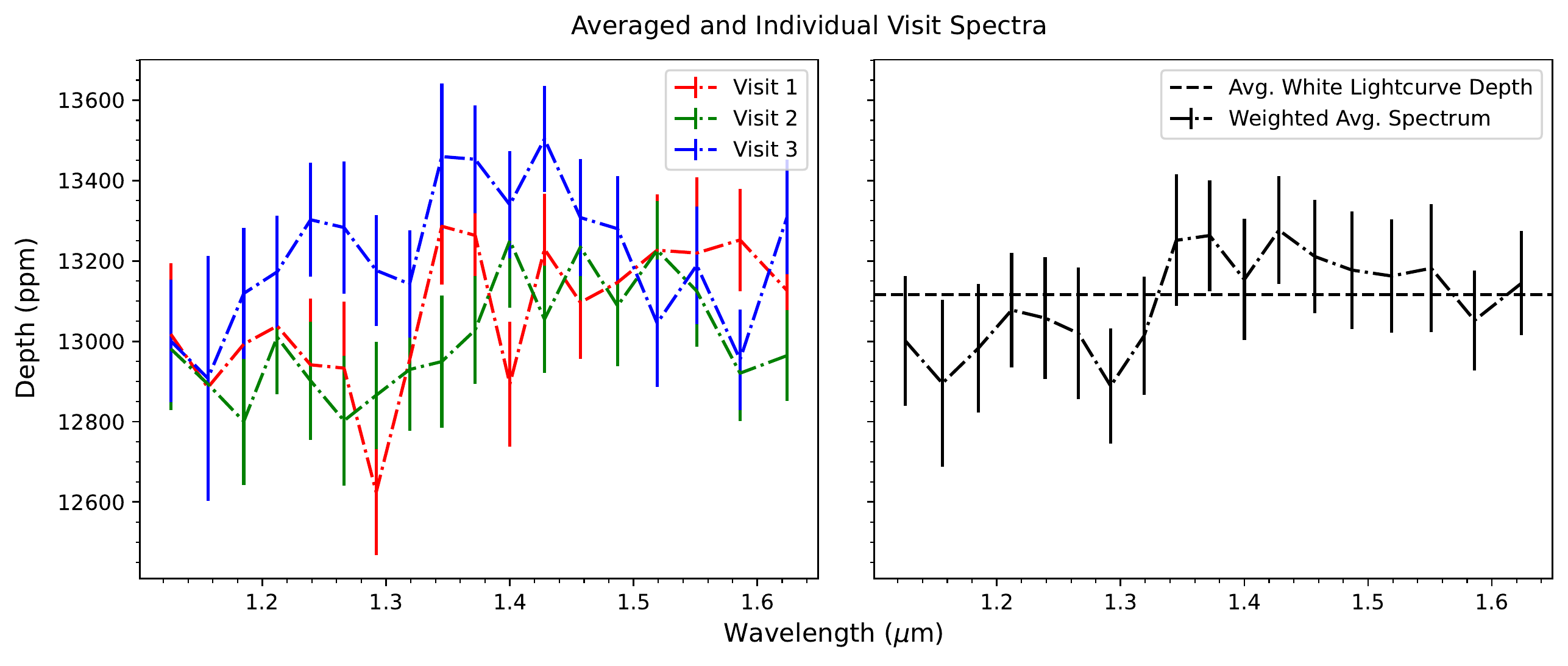}
    \vspace{-1em}
    %\caption{Unweighted average spectrum for TOI-674 b based on all three \hst{} visits, and the spectra in each individual visit.}
    \caption{The individual visit spectra for TOI-674 b, and the weighted average spectrum based on all three \hst{} visits.}
%    \vspace{-1em}
    \label{fig:spectrum}
\end{figure*}

\begin{figure*}[htb]
    \centering
    \includegraphics[width=0.8\textwidth]{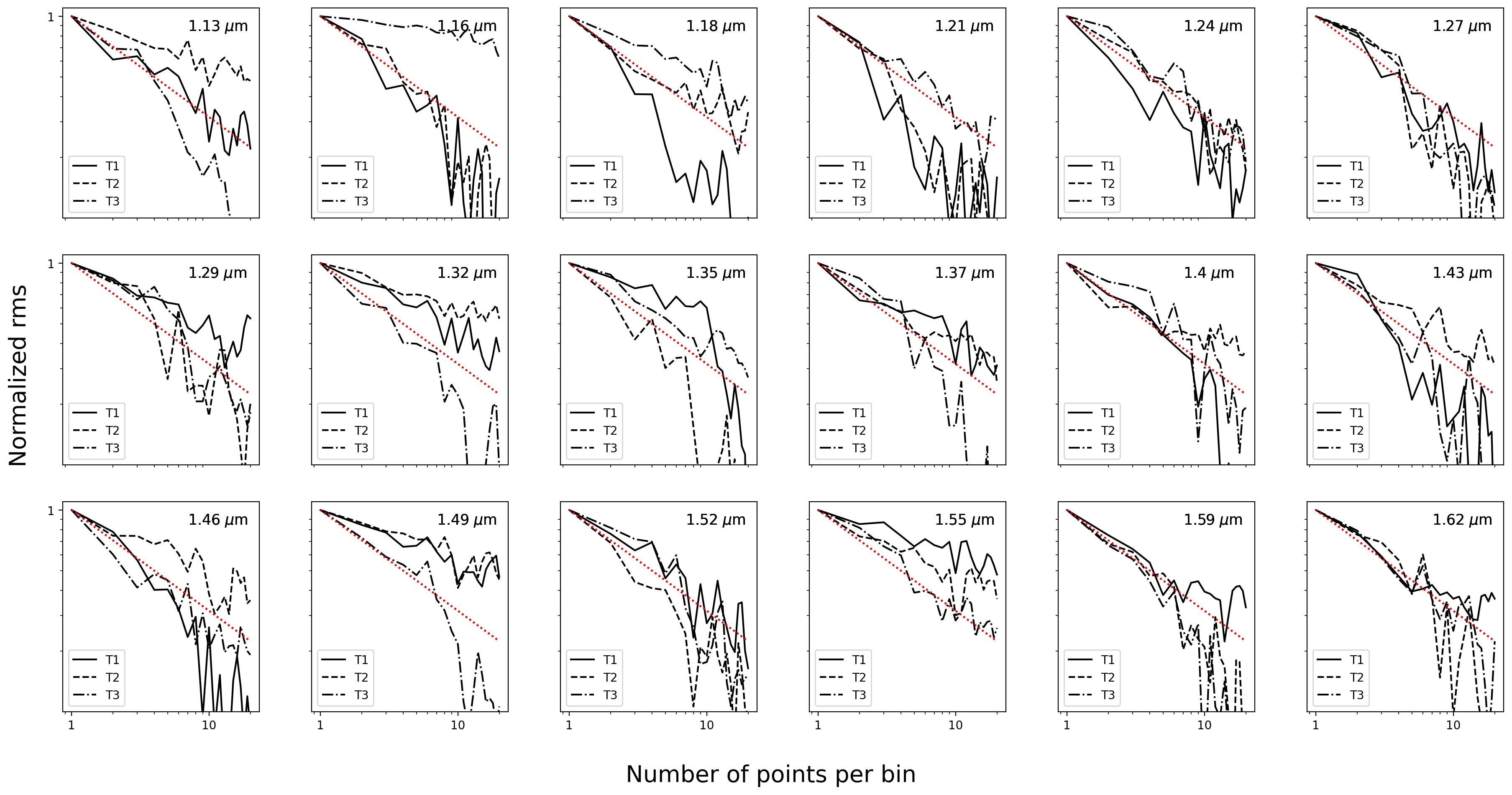}
    %\vspace{1em}
    \caption{RMS deviation plot, as a function of bin size for each observed transit of TOI-674 b. The dotted red line shows the expected $\sqrt{N}$ trend for uncorrelated noise, and the black lines show the normalized RMS trend.}
    %\vspace{1em}
    \label{fig:rms_deviation}
\end{figure*}

After modeling the WFC3 broadband transit lightcurves, we fit each of the spectral lightcurves individually, yielding the full transit spectrum. Specifically, we fixed the transit parameters to the fit white-lightcurve values except for $r_p/r_*$, and re-fit the systematics parameters with the priors described in Table \ref{tab:mcpriors}. We re-fit the systematics parameters as we noticed that they tended to be dependent on wavelength, especially in the case of the round-trip scanning flux offset. We also checked our transits for correlated noise by binning the data between 1 and 20 points and calculating the rms for each bin size. Fig. \ref{fig:rms_deviation} shows the RMS trend deviation for each spectral lightcurve bin in each visit. The expected trend due to uncorrelated noise is $\sqrt{N_{phot}}$, and our measured RMS error generally follows the uncorrelated noise trend, even though the bin sizes are relatively small. After fitting each visit, we averaged the spectra together in order to obtain the full transmission spectrum of the planet. The detrended and fitted lightcurves for transit 1 are shown in Figure \ref{fig:t1_detrend}. The detrended and fitted lightcurved for transits 2 and 3 are shown in \ref{app:a}. The measured transit depths and their uncertainties are shown in Table \ref{tab:bin_depths}. Fig. \ref{fig:spectrum} shows the final averaged spectrum from TOI-674 b, as well as the individual spectra for each \hst{} visit. Visit 3 has notably higher transit depth between $1.2~\mu$m and $1.45~\mu$m, but shows no evidence of starspot/facula crossings during the transit nor evidence of badly corrected cosmic ray hits in the data. This is seen as a consistent flux increase of $0.5\times10^8\text{ to }1\times10^8$ electrons in Visit 3 compared to Visits 1 and 2. The rotation period of the star is comparable to the interval between the second and third visit, and while stellar variability may possibly be the culprit, there is no evidence in the broadband or spectral lightcurves to indicate this in any more detail. % The flux difference is $\sim 0.5 \%$, within the observed variability seen in the full \tess{} lightcurve.}

%Visit 3 notably has higher transit depth between $1.2~\mu$m and $1.45~\mu$m, but does not significantly bias the final average spectrum. %but does not significantly change the resultant spectrum after being removed from the average.

\begin{deluxetable}{ccccc}[htb]
\tablecolumns{5}
%\tablewidth{0pt}
\tablecaption{Transit depths, transit depth errors, and limb-darkening coefficients for each spectral bin in the \tess{}, \hst{}, and \textit{Spitzer} data. \label{tab:bin_depths}}
\tablehead{
\colhead{Wavelength} & \colhead{Depth} & \colhead{Error} & \colhead{$u_1$} & \colhead{$u_2$} \\ 
\colhead{($\mu m$)} & \colhead{(ppm)} & \colhead{(ppm)} & \colhead{(fixed)} & \colhead{(fixed)}
}
\startdata
\textbf{\tess{} Depth} \\ 
0.591 -- 0.992 & 12900 & 169 & 0.098 & 0.248\\ \hline
\textbf{\hst{} Depths} \\ 
1.111 -- 1.142 & 13078 & 115 & 0.133 & 0.212\\
1.142 -- 1.171 & 12963 & 113 & 0.132 & 0.212\\
1.171 -- 1.199 & 13061 & 110 & 0.128 & 0.207\\
1.199 -- 1.226 & 13083 & 110 & 0.129 & 0.206\\
1.226 -- 1.252 & 13128 & 106 & 0.126 & 0.202\\
1.252 -- 1.279 & 13093 & 105 & 0.123 & 0.198\\
1.279 -- 1.306 & 12956 & 104 & 0.128 & 0.198\\
1.306 -- 1.332 & 13078 & 104 & 0.118 & 0.195\\
1.332 -- 1.359 & 13306 & 103 & 0.118 & 0.209\\
1.359 -- 1.386 & 13310 & 103 & 0.118 & 0.208\\
1.386 -- 1.414 & 13172 & 102 & 0.120 & 0.225\\
1.414 -- 1.442 & 13292 & 101 & 0.122 & 0.227\\
1.442 -- 1.472 & 13211 & 100 & 0.120 & 0.232\\
1.472 -- 1.503 & 13174 & 99 & 0.121 & 0.230\\
1.503 -- 1.534 & 13143 & 97 & 0.112 & 0.234\\
1.534 -- 1.568 & 13159 & 97 & 0.110 & 0.242\\
1.568 -- 1.604 & 13015 & 94 & 0.111 & 0.245\\
1.604 -- 1.643 & 13132 & 93 & 0.095 & 0.228\\ \hline
\textbf{Spitzer Depth} \\ 
3.998 -- 5.007 & 13317 & 1800 & 0.041 & 0.170\\
\enddata
\end{deluxetable}

% SHOULD I TALK ABOUT EXOPLANET AND ITS DIFFERENCES TO EMCEE/OTHER ASTRO SAMPLING PACKAGES

\subsubsection{Independent Analysis}
In addition to our method, we also used Iraclis's transmission spectroscopy modeling capabilities to conduct an independent analysis of the data. Iraclis also fits individual \hst{} visits, and uses the \texttt{divide-white} method as described in \citet{kreidberg2014}. Again, we took the unweighted average of the individual visits to obtain the transmission spectrum for our entire dataset. We found that the Iraclis results were consistent with our own analysis to within $1 \sigma$, which validates our modeling approach.

\subsubsection{Spitzer and TESS Data Points} \label{sec:spitzer}
\begin{figure*}[ht]
    \centering
    \includegraphics[width=0.6\textwidth]{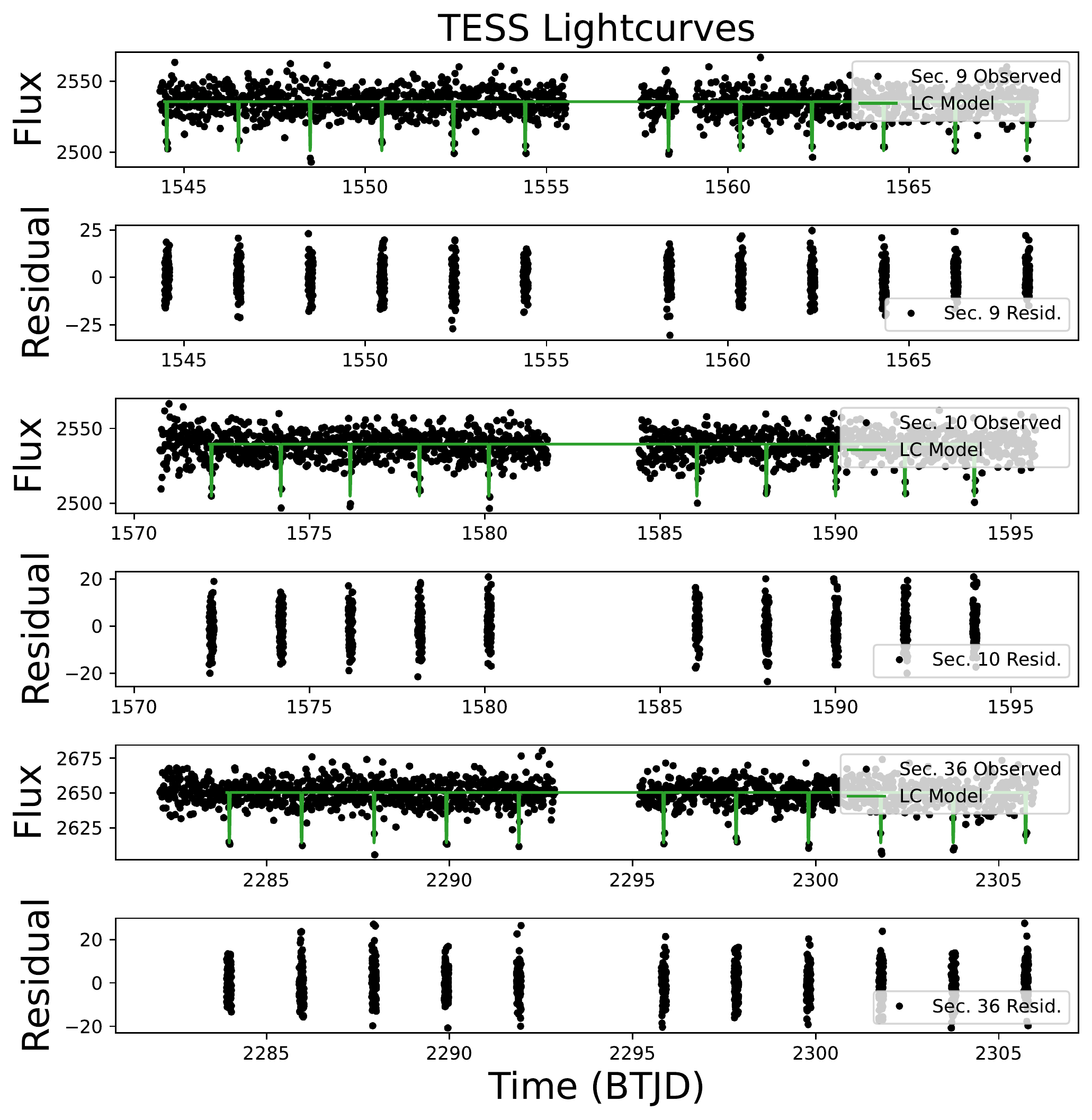}
    \caption{All three sectors of \tess{} transit data, binned to 20 min cadence, with the fitted lightcurve.}
    \label{fig:tess_lc}
\end{figure*}

\begin{figure*}[ht]
    \centering
    \includegraphics[width=0.6\textwidth]{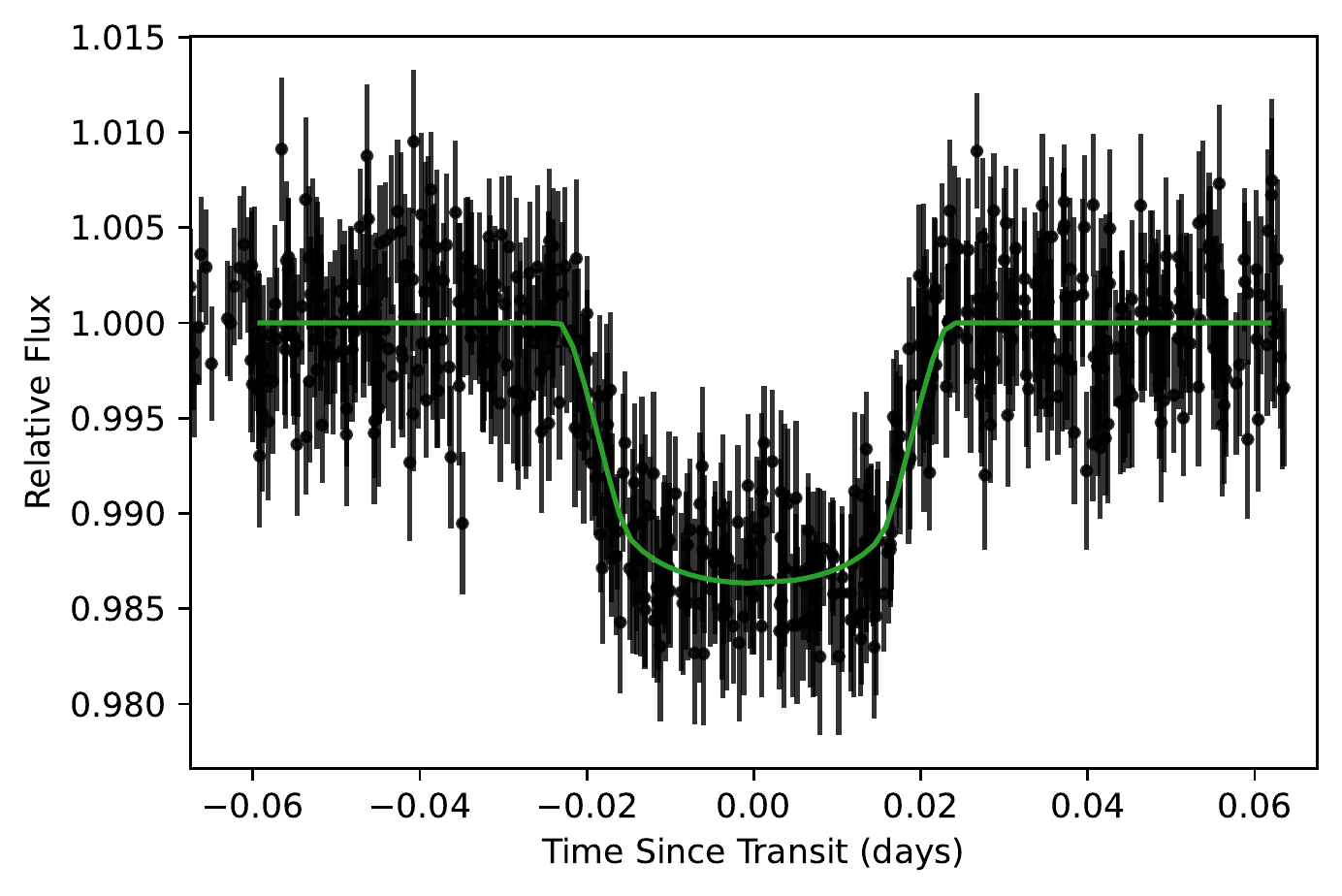}
    \caption{The folded \tess{} transit data, with the fitted lightcurve. The data has been binned to a 10-minute cadence.}
    \label{fig:tess_fold}
\end{figure*}

\begin{figure*}[htb]
    \centering
    \includegraphics[width=0.9\textwidth]{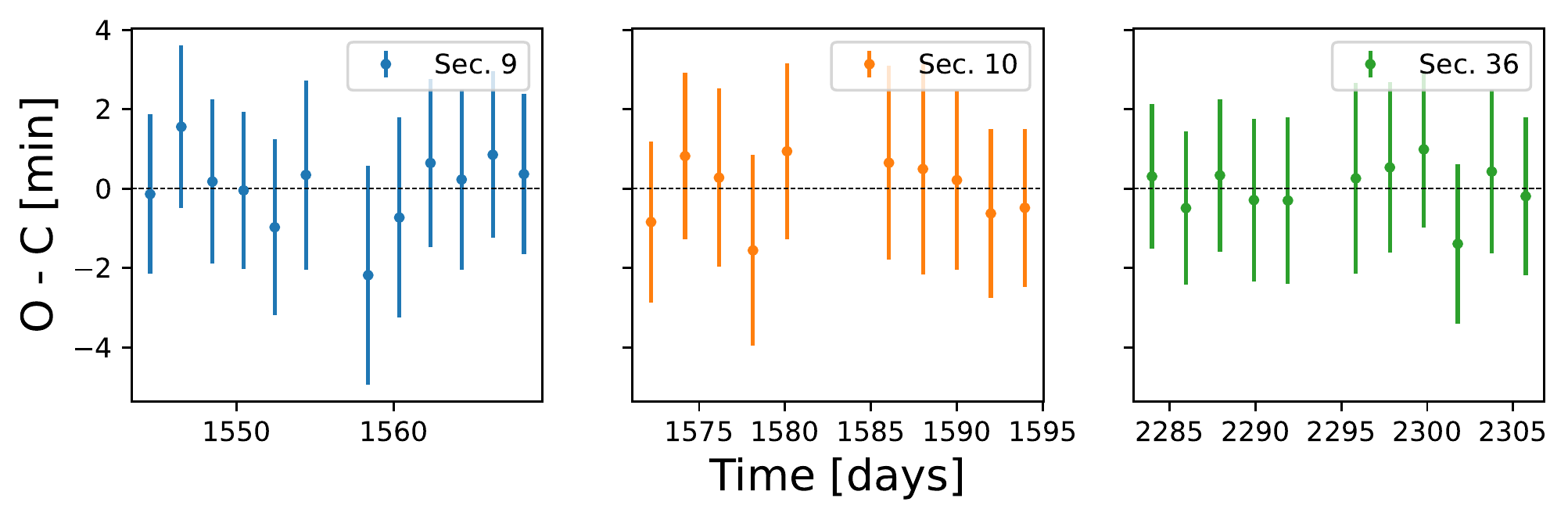}
    \vspace{-1em}
    \caption{Observed minus calculated transit times for TOI-674 b across all sectors of \tess{} data. The transit times do not significantly deviate from a linear ephemeris.}
    %\vspace{-1em}
    \label{fig:ttvs}
\end{figure*}

TOI-674 b was originally observed in \tess{} Sectors 9 and 10. The discovery paper included data from these two sectors, but TOI-674 b was also observed in \tess{} Sector 36, from 2021 March 7, to 2021 April 1. We re-fit the \tess{} data including the new sector of data in order to refine the observed and derived transit parameters, including a search for transit timing variations (TTVs) that could show evidence of undiscovered companions to TOI-674 b. Using the \texttt{exoplanet} toolkit, we fit the planet's transit parameters ($R_p/R_*$, $P$, $T_0$, $b$, $a$, $a/R_*$, $i$, and $R_p$) as well as the mean out-of-transit flux, fixing the stellar limb-darkening parameters to the values from \citet{murgas2021}, and fixing the orbital eccentricity to 0. \texttt{exoplanet} includes limb-darkened lightcurve models from \texttt{starry}, and since there were no strong systematics or stellar variability in the \tess{} lightcurve, our transit model was fairly simple:

$$M(t) = F\times M_0$$ 
\noindent
where $F$ is the out-of-transit mean flux, and $M_0$ is the bare transit lightcurve. The full fit \tess{} lightcurve is shown in Figure \ref{fig:tess_lc}, and folded transit data is shown in Fig. \ref{fig:tess_fold}. \texttt{exoplanet} also includes the ability to fit for TTVs by simply fitting the individual transit times of an otherwise Keplerian planetary orbit. By subtracting the transit time as predicted from a linear ephemeris from each fit transit time, we retrieve the TTVs of the lightcurve. More details are available in the \texttt{exoplanet} documentation\footnote{https://gallery.exoplanet.codes/tutorials/ttv/}. Including the Sector 36 data, we found $T_0 = 2458544.523792 \pm 0.000452$ BJD and $P = 1.977198 \pm 0.00007$ d. The full results of the transit analysis are shown in Table \ref{table:transit}, and the TTVs are shown in Figure \ref{fig:ttvs}. The O-C diagram shows that the transit times are consistent with a linear ephemeris, in agreement with the analysis performed in the original discovery paper. Even without a detection of a new planet in the system, the refined transit parameters and ephemerides will be useful for further studies of this planet. 

Also, as reported in the discovery paper, a single transit of TOI-674 b was observed by the \textit{Spitzer} Space Telescope on 2019 September 29 as part of a program dedicated to IRAC follow-up of \tess{} planet candidates (GO-14084, PI: Crossfield). TOI-674 b was observed at $4.5~\mu$m using \textit{Spitzer}'s IRAC instrument \citep{fazio2004}. Using the updated parameters from the full \tess{} transit fit as priors, we reanalyzed the archival \textit{Spitzer} data.  Both the \tess{} and \textit{Spitzer} transit fit results are shown in Table \ref{table:transit}. We incorporate the \tess{} and \textit{Spitzer} transit depths into our observed WFC3 spectrum for the purposes of our later atmospheric retrievals as seen in Table \ref{tab:bin_depths}.

To extract photometry from the \textit{Spitzer} observations, we used the Photometry for Orbits Eclipses and Transits (\texttt{POET} \footnote{\url{https://github.com/kevin218/POET}}) package \citep{cubillos2013, may2020}. With \texttt{POET}, we created a bad pixel mask and discarded bad pixels based on the \textit{Spitzer} Basic Calibrated Data (BCD). We ran two iterations of sigma-clipping at the $4 \sigma$ level to discard outlier pixels. We determined the center of the point spread function (PSF) is using a 2-D Gaussian fitting technique and, extracted the lightcurve using aperture photometry in combination with a BiLinearly-Interpolated Subpixel Sensitivity (BLISS) map described in \citet{stevenson2012}. We fit the resulting lightcurve with a model that accounts for both the lightcurve itself and a temporal ramp-like trend attributed to charge trapping. Finally, we sampled the posterior distributions using \texttt{POET}'s MCMC implementation with chains initialized at the best fit values.

We performed aperture photometry with various aperture sizes (ranging from 2--6 pixels in increments of 1 pixel). We found the optimal aperture size to be 3 pixels as this size returned the lowest standard deviation of the normalized residuals (SDNR). We then tested bin sizes of 0.1, 0.03, 0.01, and 0.003 square pixels for the BLISS map, and found that a bin size of 0.03~sq~px minimized the SDNR. We generated the model lightcurve using the \texttt{batman} package \citep{kreidberg2015} with $R_P/R_*$, $T_{conj}$, $a/R_*$ and $cos(i)$ as free parameters, as well as our sysematics model, a ramp-based model with a constant offset term. We held quadratic limb darkening terms constant to those found by averaging the values found by \cite{claret2011} for $T_{\text{eff}}=3500$, $\log{(g)}=5.0$, and metallicity of 0.0. We initialized four chains sampled until convergence with 10,000 burn-in steps. The resulting transit parameters are shown in Table \ref{table:transit}.

\begin{deluxetable}{lrl}
\tablecolumns{3}
\tablecaption{\tess{} and \textit{Spitzer} Transit Fits \label{table:transit}}
\tablehead{
\colhead{Parameter} & \colhead{Value} & \colhead{Error} \\
}
\startdata
\textbf{TESS Observed Parameters}\\ 
$T_0$ (BJD) & 2458544.523792 & 0.000452\\ 
P (d) & 1.977198 & 0.00007\\ 
$R_p/R_*$ & 0.1135 & 0.0006\\ 
b & 0.682 & 0.006\\ \hline
\textbf{TESS Derived Parameters}\\ 
a (AU) & 0.0231 & 0.0000003\\ 
$a/R_*$ & 11.821 & 0.0002\\ 
i (deg) & 86.69 & 0.03\\ 
$R_p$ ($R_\oplus$) & 5.20 & 0.030\\ \hline
\textbf{Spitzer Observed Parameters}\\ 
$T_0$ (BJD) & 2458756.0796 & 0.00012\\ 
$R_p/R_*$ & 0.1154 & 0.0009\\ 
b & 0.651 & 0.063\\ \hline
\textbf{Spitzer Derived Parameters}\\ 
a (AU) & 0.0243 & 0.0012\\ 
$a/R_*$ & 12.44 & 0.46\\ 
i (deg) & 87.00 & 0.27\\ 
$R_p$ ($R_\oplus$) & 5.29 & 0.17\\ \hline
\textbf{Spitzer Quad. Limb Darkening}\\ 
$u_1$ & 0.0412\\
$u_2$ & 0.170\\
\enddata
\tablecomments{Parameters without stated ranges were fixed in the transit fit.}
\end{deluxetable}

%\begin{deluxetable}{lr}
%\tablecolumns{2}
%\tablecaption{Spitzer and TESS Transit Fits \label{table:transit}}
%\tablehead{
%\colhead{Parameter} & \colhead{Value}\\ 
%}
%\startdata
%\textbf{Orbital Parameters} \\
%$P$ &  1.977164\\ 
%$T_{conj}$ &  $56.0796 \pm 0.00012$ \\ 
%$e$ &  0.0 \\ 
%$\omega$ &  0.0 \\ 
%$R_p/R_*$ &  $0.1154\pm0.0009$ \\ 
%$a/R_*$ &  $12.44\pm0.46$ \\ 
%cos$i$ &  $0.0524\pm0.0047$ \\ 
%$i$ &  $87.00\pm0.27$ \\ 
%$b$ &  $0.651\pm0.063$ \\ \hline
%\textbf{Derived Parameters} \\
%$R_p$ &  $5.29\pm0.17$ \\ 
%$a$ &  $0.0243\pm0.0012$ \\ \hline 
%\textbf{Quadratic Limb Darkening} \\
%$u1$ &  0.0412\\ 
%$u2$ &  0.170\\ \hline
%\textbf{Systematics} \\
%Constant &  $(3.3\pm 1.9)\times 10^4$\\ 
%\enddata
%\tablecomments{Parameters without stated ranges were fixed in the transit fit.}
%\end{deluxetable}

%\subsubsection{Stellar Contamination}
%Talk about if we expected stellar contamination (we didn't), and do a little bit of number crunching to say why not. \textbf{LOOK AT BERTA+ 2011}

\section{Atmospheric Retrievals}
\label{sec:retrieval}
%blah blah blah we retrieved the atmosphere of the planet from the observed spectrum

\subsection{petitRADTRANS}
We used \texttt{petitRADTRANS} (pRT), an open-source atmospheric spectral synthesis package \citep{molliere2019, molliere2020} to conduct our atmospheric retrievals. \texttt{petitRADTRANS} can be combined with several sampling packages to conduct atmospheric retrievals, and we used the suggested configuration by combining it with \texttt{PyMultiNest} \citep{buchner2014}, a Python-based implementation of the MultiNest nested sampling code \citep{feroz2009}. %pRT allows for efficient retrievals of various atmospheric configurations, and we take advantage of this capability to examine several possible atmospheres.

%We investigated two possible retrieval scenarios for our data. First, we conducted equilibrium chemistry atmospheric retrievals by fitting for the relevant planet parameters as well as the planet's metallicity (where [Fe/H]~=~0 is the solar value) and absolute carbon-to-oxygen ratio (where C/O = 0.55 is the solar value). %We also conduct retrievals where the metallicity is fixed to $100\times$ and $400\times$ solar. 
%pRT relies on precomputed tables of chemistry models in its equilibrium chemistry mode, restricting our Fe/H and C/O ranges to [Fe/H] $\in [-2, 3]$ and C/O $\in [0.1, 1.6]$. We also fit for the presence or absence of clouds, here represented as a uniform opaque gray cloud at a specific atmospheric pressure. 

In order to determine what molecules might be present in TOI-674 b's atmosphere, we conducted free chemistry retrievals for the abundances of specific atmospheric species, namely \water{}, \methane{}, \dioxide{}, and \ammonia{} \citep[all from ExoMol:][]{Chubb2021}, and CO \citep[from HITEMP:][]{rothman2010} following previous observational \citep{benneke2013}, and theoretical work \citep{miller2009}, as these are some of the dominant opacity sources in the NIR. We also fit for the presence or absence of clouds, here represented as a uniform opaque gray cloud at a specific atmospheric pressure. The full model incorporating all of the absorbers is compared to models removing one absorber at a time, and if the model without that absorber is less favored than the full model, we can say that the absorber is likely present.  %We compare models including both \water{} and \methane{} only to models including all of the possible absorbers, as \water{} and \methane{} are the most prominent absorbers in the G141 bandpass, and there is a transition from CO-dominated to \methane-dominated atmospheres around 800 K. % \textbf{(NEEDS MORE EXPLANATION, CITATION)}.
Prior distributions for our retrievals are shown in Table \ref{tab:priors}.  

\begin{deluxetable}{lcl}[htb]
\tablecolumns{3}
\tablecaption{Retrieval Priors \label{tab:priors}}
\tablehead{
\colhead{Parameter} & \hspace{1em} &\colhead{Prior} \\
}
\startdata
log(g) log$(\text{cm/s}^2)$ &\hspace{10em} & $\mathcal{U}(2.85, 3.0)$ \\
$R_{p}$ $(R_{Earth})$      &       & $\mathcal{U}(5.0, 5.5)$ \\ 
T (K)                    &       & $\mathcal{U}(600, 900)$ \\
log(P$_{cloud}$) log(bar)   &       & $\mathcal{U}(-6.0, 2.0)$ \\
\water{} log(mass frac)  &       & $\mathcal{U}(-6.0, 0.0)$ \\
\methane{} log(mass frac) &       & $\mathcal{U}(-6.0, 0.0)$ \\
CO log(mass frac) &       & $\mathcal{U}(-6.0, 0.0)$ \\
\dioxide{} log(mass frac) &       & $\mathcal{U}(-6.0, 0.0)$ \\
\ammonia{} log(mass frac) &       & $\mathcal{U}(-6.0, 0.0)$ \\ \hline
\enddata
%\tablecomments{Rows marked with * denote fixed values in the retrievals.}
\end{deluxetable}

We fixed the stellar radius, set uniform priors on the planet's radius, and temperature, and log-uniform priors on the planet's gravity, mass fraction of each absorber, and the cloudtop pressure. The retrievals were conducted using isothermal atmospheric models to create the planet spectra. % for both the equilibrium chemistry models and the free retrieval models. 
Without proper bounds on the planetary temperature, the atmospheric retrievals may find non-physical temperatures for this planet. In order to avoid this, we bound the temperature prior with some reasonable assumptions. The equilibrium temperature of a planet can be estimated either with or without incorporating heat redistribution:

$$T_{\text{eq}} = \sqrt{\frac{R_*}{a}}(1-A)^{1/4}T_{\text{eff}}$$

Or:

$$T_{\text{eq}} = \sqrt{\frac{R_*}{a}}[f(1-A)]^{1/4}T_{\text{eff}}$$

Where $f$ is a measure of heat redistribution in the range $[\frac{1}{4}, \frac{2}{3}]$ \citep{seager2010}. Without incorporating heat redistribution, and assuming a planetary albedo of 0.3, \citet{murgas2021} estimated the equilibrium temperature of the planet to be $\sim 635$ K. Assuming an albedo range of $A \in [0, 0.5]$, including the extreme bounds of heat redistribution, and for the stellar $T_{\text{eff}}=3514$ K, we calculate that the planet $T_{\text{eq}} \in [600, 900]$ K.

%Including the extreme bounds of heat redistribution (i.e. for $\text{f}=1/4$ and $\text{f}=2/3$), assuming the planet absorbs all incident radiation (i.e. $A=0$), and with the observed stellar temperature $T_{\text{eff}} = 3514$ K, we find that the expected equilibrium temperatures of the planet could range between 700 and 900 K. In order to account for a plausible range of albedos and heat redistribution efficiencies, we bound our temperature prior between 600 and 900~K.

\begin{figure*}[htb]
    \centering
    \includegraphics[width=0.9\textwidth]{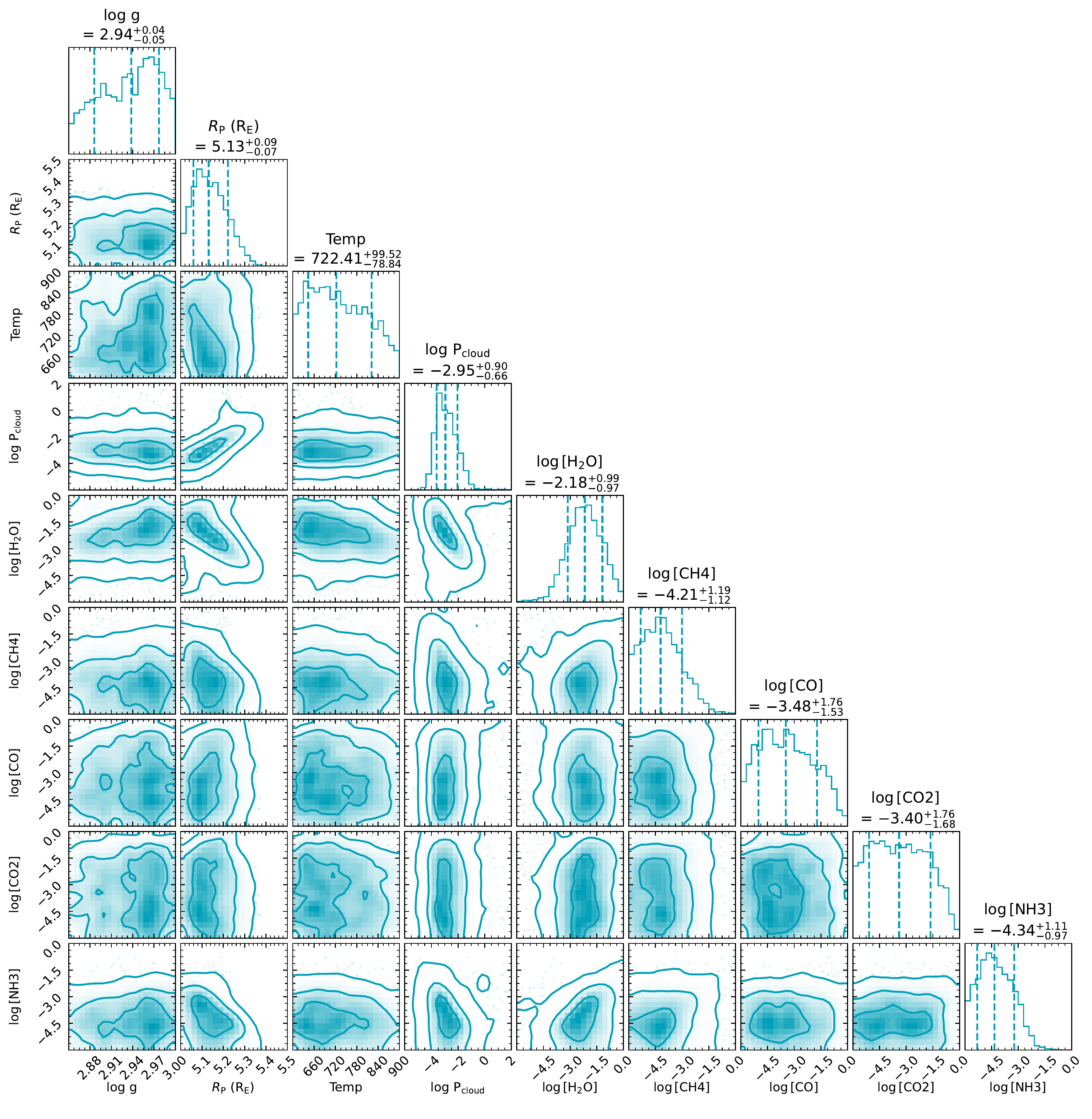}
    \vspace{-1em}
    \caption{2-D posteriors for the full retrieval model.}
    %\vspace{-1em}
    \label{fig:full_corner}
\end{figure*}

All atmospheric models were created at a resolution of 1000, and we ran Multinest to completion with 1000 live points at a sampling efficiency of 80\%. Each retrieval has an associated Bayesian evidence value $Z$, and the ratio of two evidences gives the Bayes factor $K$: 

$$K = \frac{Z_0}{Z}$$ 

where $Z_0$ is the model evidence for the full model, and $Z$ is the model evidence for a particular retrieval missing an absorber. 
Following \citet{trotta2008}, Bayes factors can be converted to p-values, and then standard deviations, by the formulas:

$$K = -\frac{1}{e (p \ln{p})}$$

where $K$ is the Bayes factor and $p$ the p-value, and:

$$p = 1 - \text{erf}\left(\frac{n_{\sigma}}{\sqrt{2}}\right)$$

where $n_\sigma$ is the sigma significance and erf is the error function. \citet{trotta2008} and \citet{benneke2013} present ranges of Bayes factors that correspond to p-values and sigma significances, with $2.9 \leq K < 12$ ($2.1\sigma \leq n_\sigma < 2.7\sigma$) a ``weak detection", $12 \leq K < 150$ ($2.7\sigma \leq n_\sigma < 3.6\sigma$) a ``moderate detection", and $K \geq 150$ ($n_\sigma \geq 3.6\sigma$) a ``strong detection". The Bayes factor analysis results are shown in Table \ref{tab:evidence}. We find that the presence of \water{} is moderately favored with a Bayes factor of 15.8, corresponding to a $2.9\sigma$ detection, and we find weak evidence for the presence of clouds at a Bayes factor of 4.0 ($2.2\sigma$), but the evidence for the other absorbers is insignificant. We also present the best-fit values for the full model in Table \ref{tab:bf_results}, and the 2-D posterior distributions in Figure \ref{fig:full_corner}.

%Bayes factors can be used to evaluate the strength of some particular evidence by converting them  %according to the following ranges \citep{kass1995}: $K < 3.2$ insignificant, $3.2 \leq K < 10$ substantial, $10 \leq K < 100$ strong, and $100 \leq K$ decisive.

\subsubsection{Utility of Equilibrium Chemistry Models}
Previous theoretical \citep{moses2013} work on Neptune-sized exoplanets has revealed a diversity of potential atmospheric compositions ranging from ``typical" hydrogen/helium dominated atmospheres to ``exotic" atmospheres dominated by significantly heavier gases such as \dioxide{} and \water{}. Observationally, warm Neptune atmospheres range from clear \citep[HAT-P-11 b:][]{fraine2014} to cloudy \citep[GJ 436 b:][]{knutson2014a}, and low-metallicity \citep[HAT-P-26 b:][]{benneke2019} to high-metallicity \citep[GJ 3470 b and GJ 436 b:][]{wakeford2017, morley2017}. %even for the handful for which there are measured transmission spectra (e.g. HAT-P-11 b \citep{fraine2014}, HAT-P-26 b \citep{wakeford2017}, GJ 436 b \citep{knutson2014b}, and GJ 3470 b \citep{benneke2019}). Interestingly, these planets have been observed to have very diverse metallicities, ranging from near solar \citep{benneke2019} to high metallicity \citep{morley2017}, suggesting varied planet formation processes. 
We compared our observed spectra to self-consistent atmospheric models (interpolated chemical abundances based on nonequilibrium chemistry in the \texttt{easyCHEM} grid described in \citet{molliere2017}) and found that our data are not precise enough, nor do they have the wavelength coverage to distinguish between high-metallicity clear models and low-metallicity cloudy models (as seen in Figure \ref{fig:jwst_example}). Future observations with JWST would provide both better precision as well as sensitivity across wider bandpasses, allowing future investigators to study the equilibrium chemistry of this planet further.

%Ordinarily it would be desirable to use equilibrium chemistry models to fit for the metallicity of the planet and then determine the abundances of atmospheric species, but these predictions may not be consistent with observations due to nonequilibrium processes (see e.g. \citet{fortney2020} for a detailed treatment of the varied observable tracers) or unknown formation history \citep{benneke2019}. Given these theoretical and observational uncertainties, we do not attempt to constrain the atmospheric metallicity of TOI-674 b.

%The table shows that the atmospheric models we investigated are hard to distinguish, although the 400x solar metallicity cloudy model has marginally substantial preference over the base case. We show the best fit spectrum and corner plot for this configuration in Fig. \ref{fig:retrieval_spectra} and Fig. \ref{fig:retrieval_corner}.

\begin{deluxetable*}{lrrrrrrr}[htb]
\tablecolumns{7}
\tablecaption{Bayesian Evidences for Various Retrieval Scenarios \label{tab:evidence}}
\tablehead{
\colhead{Retrieval Model} & \colhead{DOF} & \colhead{$\chi^2$} & \colhead{$\chi_\nu^2$} & \colhead{BIC} & \colhead{$\log_{10}(Z)$} & \colhead{$\Delta\log_{10}(Z)$} & \colhead{Bayes Factor for} \\
\colhead{} & \colhead{} & \colhead{} & \colhead{} & \colhead{} & \colhead{} & \colhead{} & \colhead{molecule present} \\
}
\startdata
\textbf{Full Model} & & & & & & & \\
H2O, CH4, CO, CO2, NH3, Cloudy & 11 & 7.5 & 0.7 & 34.4 & -5.5 & 0.0 & 1.0 \\ \hline
\textbf{No H2O} & \textbf{12} & \textbf{15.1} & \textbf{1.3} & \textbf{39.1} & \textbf{-6.7} & \textbf{1.2} & \textbf{15.8} \\
No Cloud & 12 & 9.0 & 0.8 & 33.0 & -6.1 & 0.6 & 4.0 \\
No CO2 & 12 & 6.5 & 0.5 & 30.5 & -5.3 & -0.2 & 0.6 \\
No NH3 & 12 & 8.1 & 0.7 & 32.1 & -5.3 & -0.2 & 0.6 \\
No CO & 12 & 7.3 & 0.6 & 31.3 & -5.3 & -0.2 & 0.6 \\
No CH4 & 12 & 6.7 & 0.6 & 30.7 & -5.2 & -0.3 & 0.5 \\ \hline
Featureless & 16 & 18.8 & 1.6 & 42.8 & -6.1 & 0.6 & 4.0 \\
Constant-Depth & 19 & 20.1 & 1.1 & 23.1 & N/A & N/A & N/A \\
Linear & 18 & 17.6 & 1.0 & 23.5 & N/A & N/A & N/A \\
\enddata
%\tablecomments{$\Delta \log_{10}$(Z) and Bayes Factors calculated relative to the full model}
\end{deluxetable*}

\begin{figure*}[htb]
    \centering
    \includegraphics[width=0.8\textwidth]{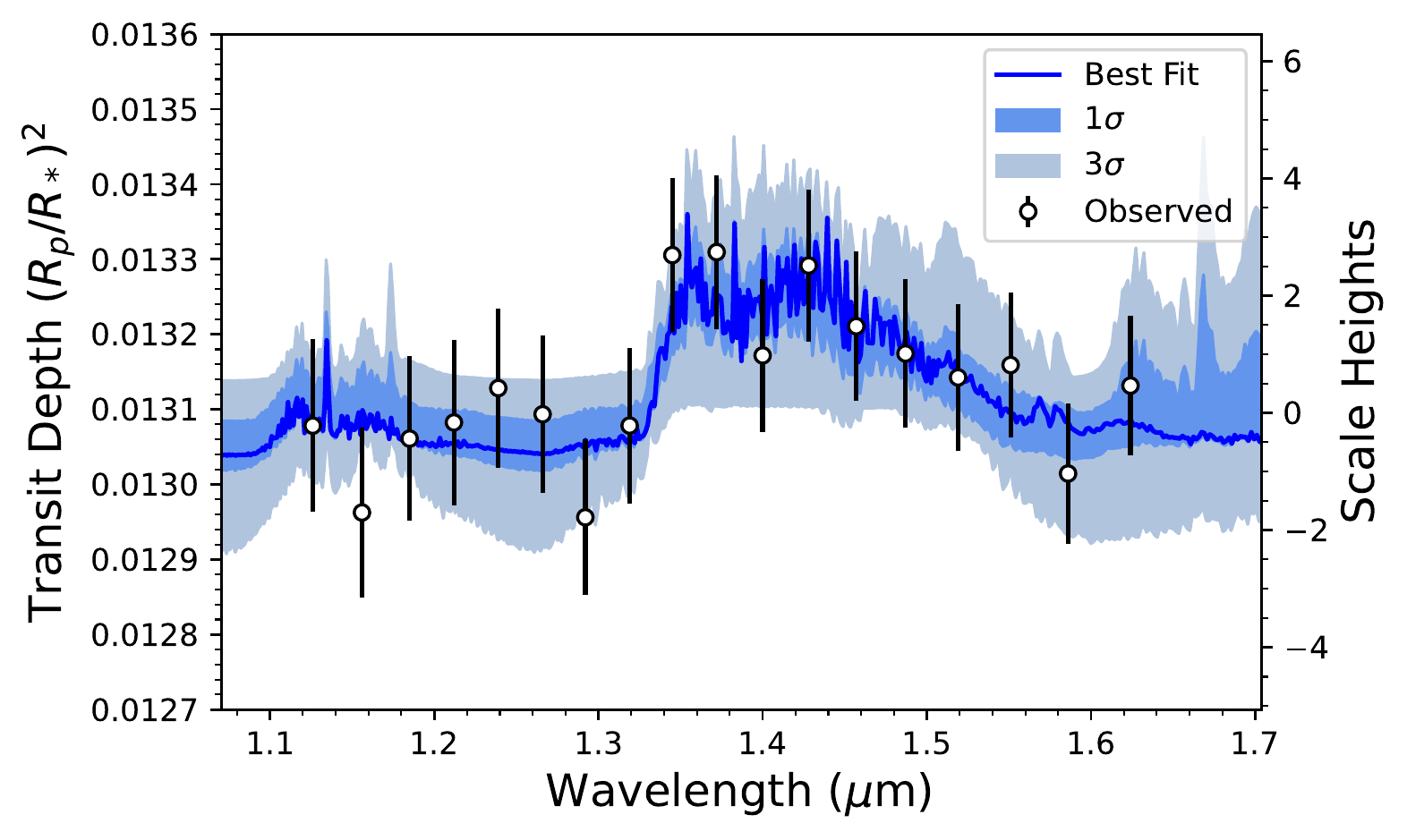}
    \vspace{-1em}
    \caption{3- and 1-sigma intervals for the full model. The solid blue line is the best fit model from the full retrieval.}
    %\vspace{-1em}
    \label{fig:h2o_sampled}
\end{figure*}

%\begin{figure*}[htb]
%    \centering
%    \includegraphics[width=0.8\textwidth]{posterior_simple.pdf}
%    \vspace{-1em}
%    \caption{2-D posteriors for the water abundance and cloudtop pressure. Cloudtop pressure is given in $\log_{10}$(bar), and water abundance is given in $\log_{10}$(mass mixing ratio).}
    %\vspace{-1em}
%    \label{fig:h2o_corner}
%\end{figure*}

\begin{deluxetable*}{lr}[htb]
\tablecolumns{2}
\tablecaption{Full Atmospheric Model Results \label{tab:bf_results}}
\tablehead{
\colhead{Parameter} & \colhead{Fit Value} \\
}
\startdata
%\textbf{Full Retrieval} \\ \hline
%*$R_* ({R_{Sun}})$              & 0.42\\
log(g) $(\text{cm/s}^2)$        & $2.94 \pm 0.05$ \\
$R_{pl} (R_{Earth})$            & $5.13 \pm 0.08$ \\ 
T (K)                           & $723 \pm 90$ \\
H2O                       & $-2.2 \pm 1.0$ \\
CH4                       & $-4.2 \pm 1.2$ \\
CO                        & $-3.5 \pm 1.7$ \\
CO2                       & $-3.4 \pm 1.7$ \\
NH3                       & $-4.3 \pm 1.0$ \\
log(P$_{cloud}$) (bar)          & $-2.94 \pm 0.8$  \\ \hline
\enddata
\tablecomments{pRT abundances are given as $\log_{10}$(mass mixing ratio). These can be converted to volume mixing ratios by $n_i = X_i \frac{\mu}{\mu_i}$, where $n_i$ is the VMR, $X_i$ the mass fraction of a species, $\mu_i$ the molecular weight of the species, and $\mu$ the mean molecular weight of the atmosphere.}
\end{deluxetable*}

%The Bayesian evidence tables shows that no atmospheric model is strongly favored compared to the base cases. However, the cloud 400x Solar metallicity equilibrium chemistry model and the cloudy H$_2$O/CH$_4$ free retrieval model are marginally favored over their base cases. We show the best fit spectra and corner plots for these models in Figs. \ref{fig:400x_spec}, \ref{fig:400x_corner}, \ref{fig:h2o_spec}, \ref{fig:h2o_sampled}, and \ref{fig:h2o_corner}. We also present the best-fit values for the 400x Solar, Cloudy Eq. Chem. model and the Cloudy H2O/CH4 model in Table \ref{tab:bf_results} in Appendix \ref{app:a}.

%Given the best-fit parameters from our equilibrium chemistry retrievals, we can compute the abundances of interesting species in the atmosphere of TOI-674 b. \textbf{IS THIS ACTUALLY REASONABLE TO DO? DO WE TRUST WHATEVER WE'D GET OUT OF THIS?}

\section{Discussion} \label{sec:discussion}
\subsection{Atmospheric Compositions}

\begin{figure*}[htb!]
    \centering
    \includegraphics[width=0.8\textwidth]{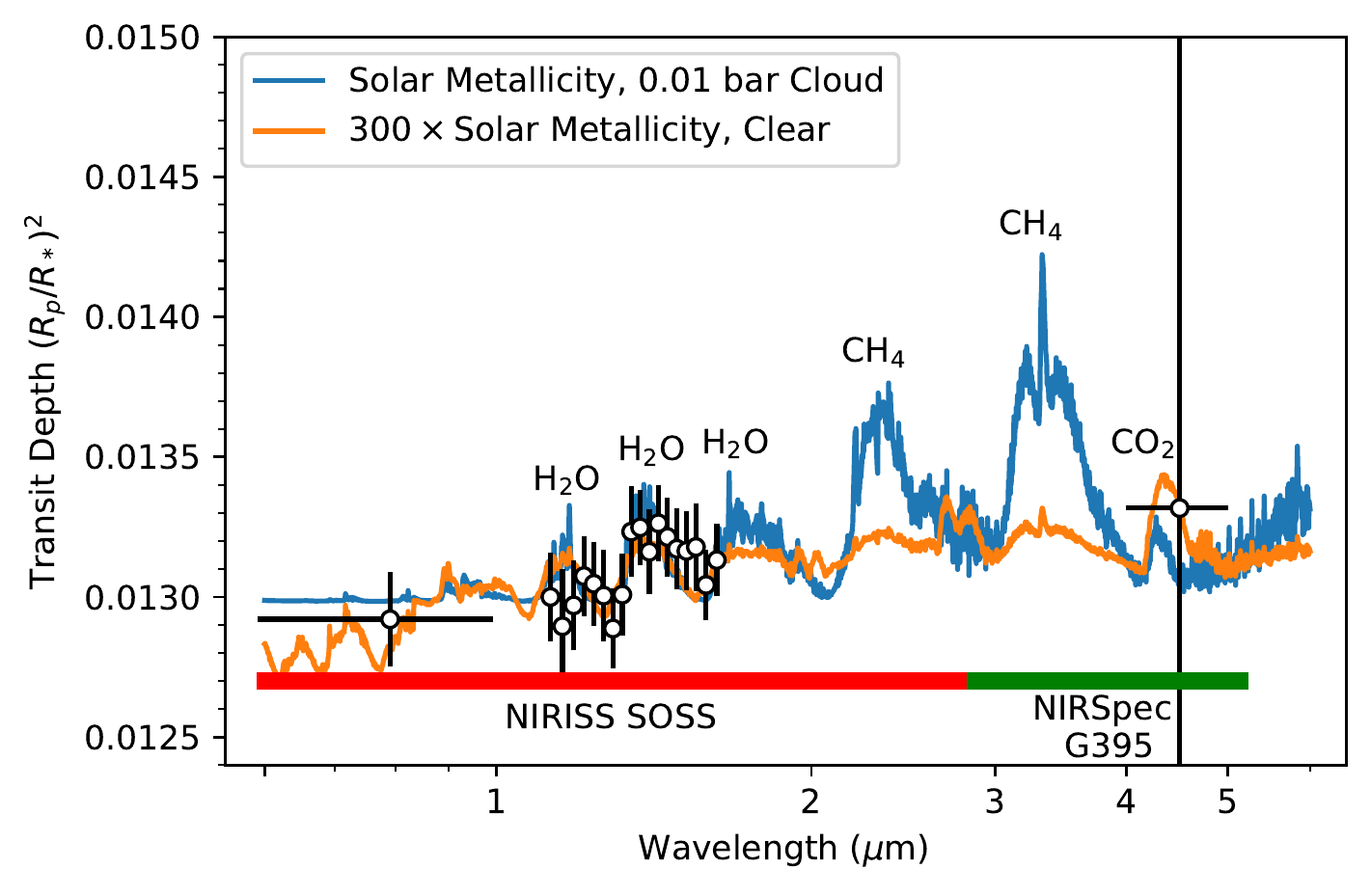}
    \vspace{-1em}
    \caption{Example \textit{JWST} Cases: one with solar metallicity and high altitude clouds, and one with high metallicity and a clear atmosphere. Both are consistent with our observed \hst{} WFC3 data, and need the precision that JWST provides to disambiguate the two models and precisely measure the abundances of the marked absorbers.}
    %\vspace{-1em}
    \label{fig:jwst_example}
\end{figure*}

Although our retrieval analysis is useful for identifying the presence of particular absorbers, our data are not precise enough to allow us to precisely measure the abundances of any absorbers present. In this case, a range of \water{} abundances is likely to be consistent with the data, as seen in Table \ref{tab:bf_results}. Each absorber has at least an order of magnitude uncertainty in the mass fraction, and some (like \dioxide{} and CO), have error bars of close to two orders of magnitude. As for the goodness-of-fit of our models, in only a single case, the preferred no \water{} model, is the reduced chi-square value close to and greater than 1. While this may indicate over-fitting on the part of our atmospheric models, or we are still over-estimating the uncertainties on our transit depths. Better quality data, perhaps from \textit{JWST}, would allow for more precise transit obervations with fewer systematic effects to correct. Assuming a mean molecular weight $\mu = 3.0$ amu (corresponding to $\sim30\times$ Solar metallicity), we estimate a scale height $\text{H}\sim260$ km, approximately equal to 100 ppm transit depth per scale height. The amplitude of the $1.4\mu$m water feature here is $\sim 2$ scale heights, somewhat higher than expected from the trend in \citet{crossfield2017} given the range of possible equilibrium temperatures for TOI-674~b.  Further work will explore this trend in more detail including an updated sample of Neptune-sized exoplanets with measured transmission spectra. The prominence of these features is likely to be dependent on both cloudtop pressure and atmospheric metallicity. For example, both a solar metallicity atmosphere with a 0.01 bar cloud and a $300\times$ solar metallicity clear atmosphere are consistent with our observed \hst{} data. A significant diversity of atmospheric metallicities are predicted from formation modeling, from very high metallicities \citep{fortney2013}, to very low \citep{bitsch2021}, depending on where and how the planet formed in its disk, and whether it migrated relative to the the frost lines. Higher resolution, higher photometric precision data from a larger telescope will be critical to constraining TOI-674 b's atmospheric metallicity to inform planetary formation models. The recently launched James Webb Space Telescope will be able to acquire much better quality data across a larger NIR bandpass than can currently be collected by \hst{}, allowing access to distinct \water{}, \methane{}, and \dioxide{} features across the NIRISS, NIRSpec, and MIRI bandpasses (see Fig. \ref{fig:jwst_example}, and \citet{greene2016} for an observability study). \dioxide{} in particular is a tempting molecule to detect, as it is a better tracer of atmospheric metallicity than \water{} \citep{moses2013}.

%\subsection{Potential Stellar Contamination}
%Possibly ask Ben Rackham what a good rule of thumb is for stellar contamination amplitudes. Tom Barclay suggested a 200ppm feature is likely to dwarf any potential contamination signal

%\subsection{Alternate Retrievals}
%put whatever Bjorn does here

\subsection{Possible Helium Escape Observations}
In addition to future space-based near- and mid-infrared transmission spectroscopy, there is also room to further characterize TOI-674 b and its relatively unique place as an M-dwarf planet in the Neptune desert. As a low-mass Neptune desert planet, TOI-674 b is likely to be undergoing or have undergone potentially significant atmospheric escape due to stellar irradiation. One such tracer for this evolutionary process is the metastable helium line at 10830 \AA{} \citep{oklo2018}. The WFC3 G102 grism can measure the potential metastable helium transit of TOI-674 b from space \citep[WASP-107~b;][]{spake2018}, and metastable helium exospheres have also been observed with ground-based high-resolution spectrographs \citep[HAT-P-11~b;][]{allart2018}. With this in mind, we simulated the expected helium absorption signature.

\subsubsection{Atmospheric Simulations}

To estimate the expected absorption signature in the 10830 \AA{} line triplet of neutral helium, we simulate the atmosphere of TOI-674 b using a spherically symmetric atmospheric escape model \citep{oklo2018}. The density and velocity profiles of the escaping atmosphere are based on the isothermal Parker wind \citep{parker1958, lamers1999} and the model atmosphere is composed entirely of atomic hydrogen and helium, with a 9:1 number ratio. The main free parameters are the temperature of the upper atmosphere %(thermosphere) 
and the total mass loss rate, but without information on the high-energy luminosity of the host star, it is difficult to constrain their values. If we assume that the stellar spectrum is similar to that of GJ 176, an M2.5-type star observed as part of the MUSCLES survey \citep{france2016}, the energy-limited mass-loss rate would be on the order of $10^{10}$~g~s$^{-1}$. 

\begin{figure}[t]
    \centering
    \includegraphics[width=0.5\textwidth]{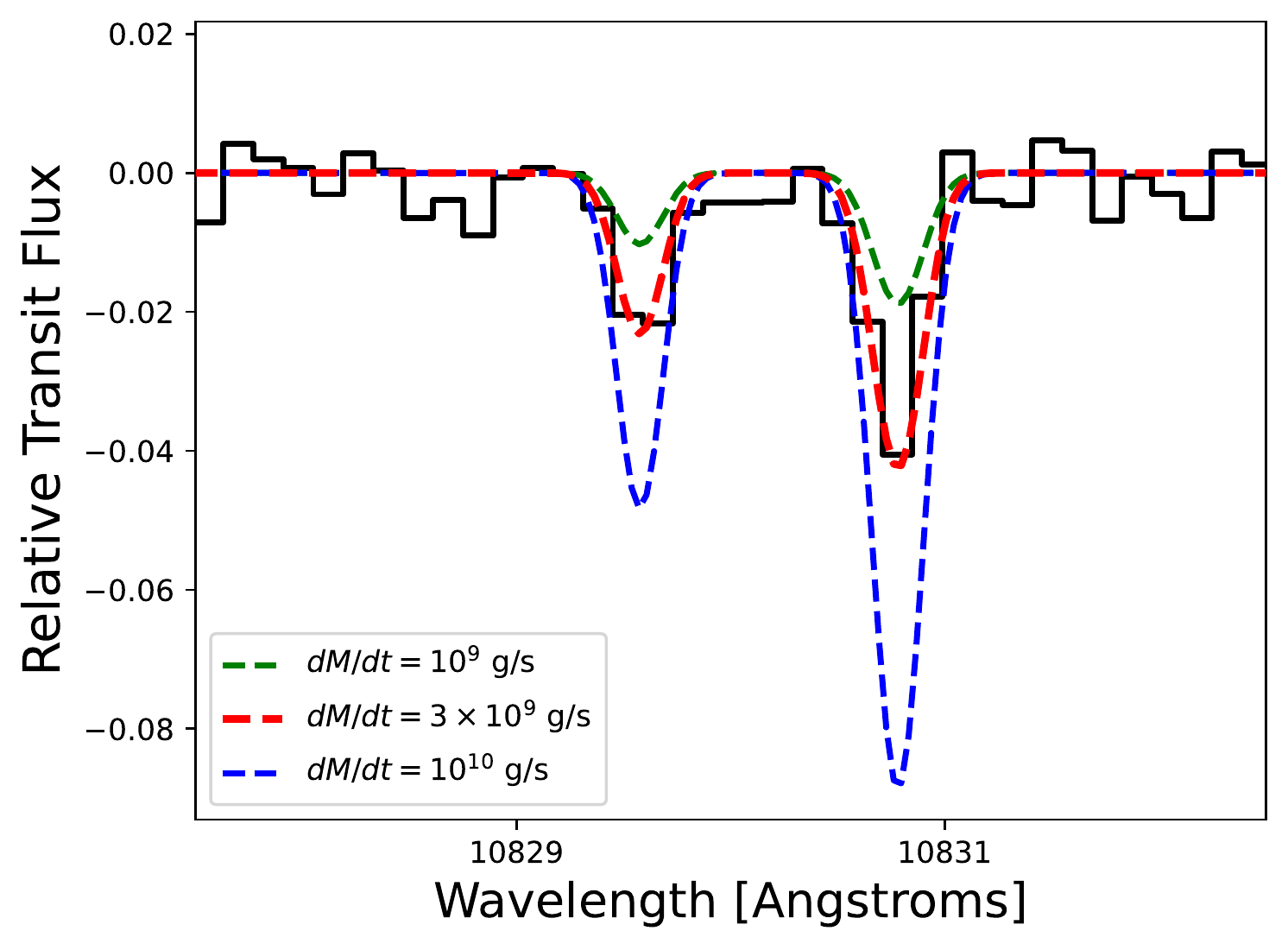}
    \vspace{-1em}
    \caption{Simulated helium signatures for TOI-674 b, assuming a 4000K thermosphere, showing how the strength of the signal varies with mass-loss rate. We also show the expected performance of a 10-meter class observatory observing a single transit of TOI-674~b, for the $3\times10^9$ g/s mass-loss rate, detecting the helium signature at a S/N of 7.}
    %\vspace{-1em}
    \label{fig:helium}
\end{figure}

We ran a grid of models spanning a range of thermosphere temperatures between 4,000 K and 9,000 K, and mass-loss rates between $10^9$~g~s$^{-1}$ and $10^{11.5}$~g~s$^{-1}$. 
We note that in planets undergoing helium escape, most of the helium opacity comes from $\sim1.5$--$3 R_p$. We also note that in \citet{salz2016}, the corresponding thermosphere temperatures for similar low-gravity gaseous planets (GJ 3470 b and GJ 436 b) at these planetary radii range from $\sim4000$--$9000$K, giving the thermospheric temperature range. We perform radiative transfer calculations along the planet's terminator, using the MUSCLES spectrum of GJ 176 as input, in order to calculate the abundance of helium atoms in the excited 2$^3$S state and the resulting opacity at 10830~\AA{}. Finally, we compute the transmission spectrum for the planet at mid-transit. The predicted excess absorption depths vary substantially depending on the assumed model parameters (as shown in Fig. \ref{fig:helium}), but in many cases the level of absorption is on the order of several percent at the line center, making this planet potentially interesting for helium 10830~\AA{} observations.

\section{Conclusions}
We present the \hst{} WFC3 G141 infrared transmission spectrum of the warm Neptune TOI-674~b. We reduced the WFC3 data and extracted the spectral lightcurves with \texttt{Iraclis}, and detrended and fit the lightcurves with \texttt{exoplanet}.  We also re-fit the \tess{} data including new observations from \tess{} Sector 36 with \texttt{exoplanet} in order to update the planet's transit parameters, and re-fit the archival \textit{Spitzer} $4.5\mu$m photometry with POET. We also searched for and found no evidence of transit-timing variations in the planet's \tess{} lightcurve. Both the \tess{} and \textit{Spitzer} transit depths were incorporated into the planet's observed transmission spectrum. After conducting atmospheric retrievals on the observed transmission spectrum with petitRADTRANS, we find moderate evidence ($2.9\sigma$) for increased absorption in the atmosphere of the warm Neptune TOI-674~b due to water vapor, and weak evidence (2.2$\sigma$) for the presence of clouds.

%TOI-674 b joins the club of planets with measured transmission spectra, and the even more exclusive club of planets for which those spectra are not flat. 
Other than TOI-674 b, only three other Neptune-size planets (masses between 10 and 40 Earth masses) have notable features in their atmospheres (WASP-107 b: \citet{kreidberg2018, spake2018}, HAT-P-11 b: \citet{chachan2019}, and HAT-P-26 b: \citet{wakeford2017}).  With water present in its atmosphere, TOI-674~b is a good candidate for further study to determine the other components of its atmosphere, as well as potential tracers of atmospheric mass loss. Future work should concentrate on these efforts, especially as \tess{} continues to discover these types of exoplanets around nearby stars. Only by characterizing a large sample of Neptune-like exoplanets will we be able to more fully understand the formation and migratory processes that lead to the observed diverse population of exo-Neptune orbital architectures.

\begin{acknowledgments}
This work was conducted on the ancestral territory of the Kaw, Osage, and Shawnee peoples. We thank Paul Mollière and Evert Nasedkin for their extremely helpful assistance with petitRADTRANS. This work was supported in part by a grant from the NASA Interdisciplinary Consortia for Astrobiology Research (ICAR). This research made use of the open source Python package ExoCTK, the Exoplanet Characterization Toolkit \citep{bourque2021}. 

% Thanks to Paul Molliere and Evert Nasedkin for pRT questions
% This work was conducted on the tribal homelands of the Kaw, Kickapoo, Sioux, Osage, and Shawnee peoples. ex: https://ceop.ku.edu/land-acknowledgement
\end{acknowledgments}

\appendix
\section{Fitting the Round-trip Spatial Scan Systematic} \label{app:scan}

Round-trip spatially-scanned WFC3 IR data has a significant flux offset as the target star is scanned up or down the detector. This flux offset is due to the effect of the motion of the spatial scan combining with the direction of the detector readouts. If the scan is proceeding in the same direction as the detector's row-by-row readout, the effective exposure time will be greater than if the scan is proceeding in the opposite direction as the readout. Thus, electron counts will be higher for the downstream scans than the upstream scans. \citep{mccullough2012}. Historically, (e.g. \citet{knutson2014b}, for HD 97658 b) the upstream and downstream scans have had lightcurves extracted, detrended, and fit independently and only later combined to find the true transit model. We believe the sinusoidal approach is more efficient by allowing both scan directions to be fit in the same operation as opposed to fitting both scans separately. However, in order to demonstrate that our approach is valid, we compare it to the legacy method.

\begin{figure*}[htb]
    \centering
    \includegraphics[width=0.6\textwidth]{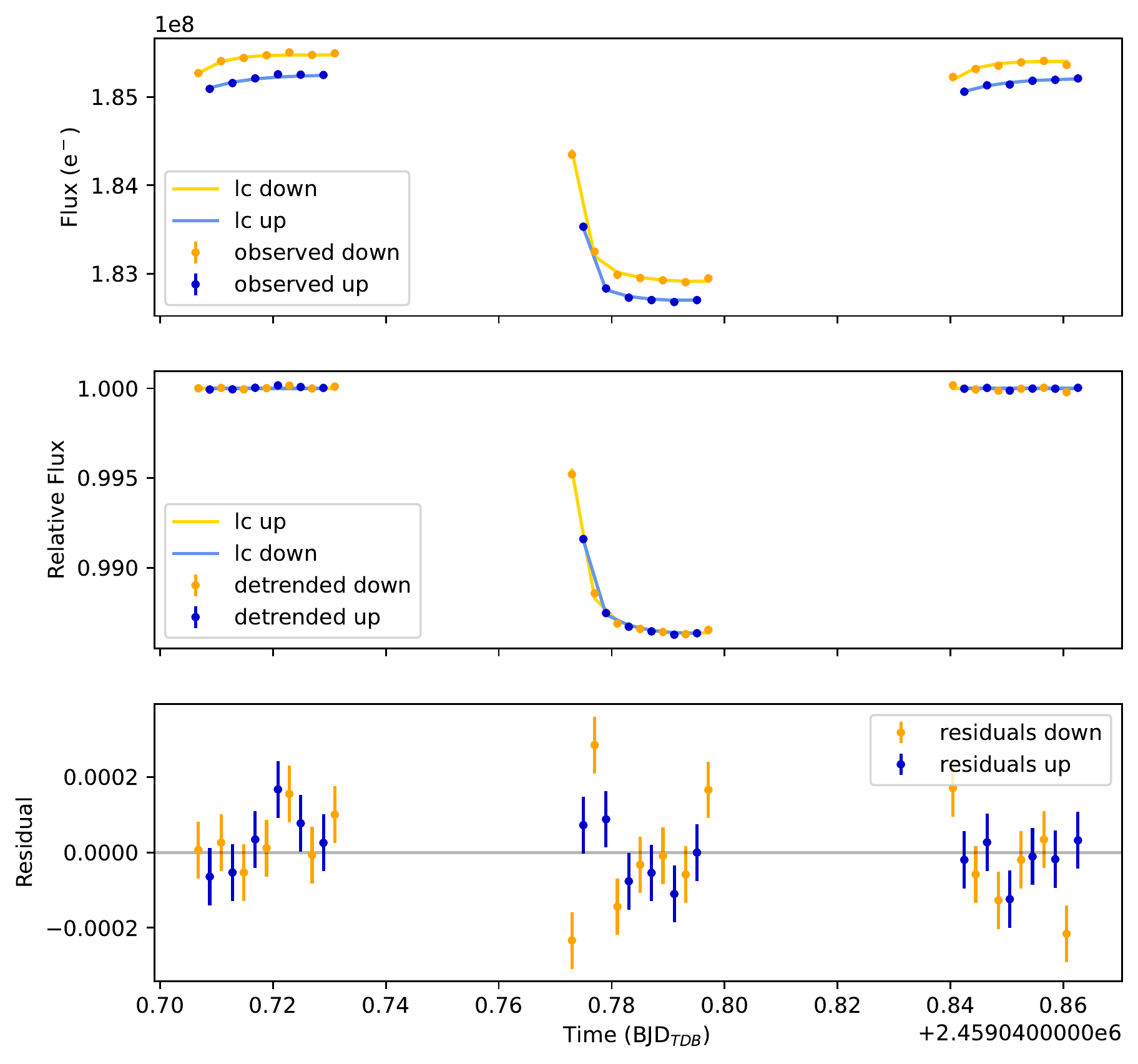}
    \vspace{-1em}
    \caption{White lightcurve fits for modeling the systematics in each scan direction independently.}
    %\vspace{-1em}
    \label{fig:scan_white}
\end{figure*}

We re-fit the white lightcurve for Visit 1 of TOI-674 b according to the transit and systematics model provided in \citet{knutson2014b}: 
$$F(t) = c_1 (1 + c_2 t + c_3 e^{-p/c4}) F_{transit}(t)$$ 
where $c_1$--$c_4$ are free parameters ($c_1$ the out of transit median flux, $c_2$ the visit-long slope, $c_3$ the ramp amplitude, and $c_4$ the ramp timescale), $t$ the time in days, and $p$ the time in days since the first exposure in the visit. $F(t)$ is the full systematics-included transit lightcurve model, and $F_{transit}(t)$ is the transit-only model. The planet's transit parameters were shared across both scans, and the systematics parameters were fit independently for each scan direction, using the same sampler configuration as the main analysis in this work. The combined transit and systematics model fits are shown in Figure \ref{fig:scan_white}.

After fitting the Visit 1 white lightcurve scan directions separately, we compared the found planet transit parameters ($T_0$ and $R_p/R_*$) to our main analysis. %As the systematics models were not directly comparable, we present the relevant transit parameters, $T_0$ and $R_p/R_*$. 
The values closely agree, easily within $1\sigma$, as seen in Table \ref{table:scan_compare}. In addition, the legacy separate-scan fit method had 17 total parameters and took 2m 22s to run, while our sinusoid method had 13 parameters and took 1m 16s to run. In order to more directly compare the two methods, we calculated the Bayesian Information Criteria (BIC) for the two models, where the model with the lower BIC is preferred. The separate-scan method had 21 degrees of freedom, a BIC of 135.4, and an RMS error of $1\times10^{-4}$, while our sinusoid method had 25 degrees of freedom, a BIC of 127.6 and an RMS error of $1.1\times10^{-4}$. Given the close agreement of the transit parameters between the models, and that our sinusoid method has a lower BIC than the separate-scan method, we are confident that our method is equivalent to or better than fitting the scan directions separately.

\begin{deluxetable}{lrr}
\tablecolumns{3}
\tablecaption{Separate-Scan vs. Sinusoid Transit Parameters \label{table:scan_compare}}
\tablehead{
\colhead{Parameter} & \colhead{Separate-Scan} & \colhead{Sinusoid}\\
}
\startdata
$T_0$ (BJD) & 2459040.79430 $\pm$ 0.00004 & 2459040.79429 $\pm$ 0.00004 \\ 
$R_p/R_*$ & 0.1144 $\pm$ 0.0002 & 0.1144 $\pm$ 0.0002\\ 
\enddata
%\tablecomments{Parameters without stated ranges were fixed in the transit fit.}
\end{deluxetable}

\section{Additional Plots} \label{app:a}
\begin{figure*}[htb]
    \centering
    \includegraphics[width=0.6\textwidth]{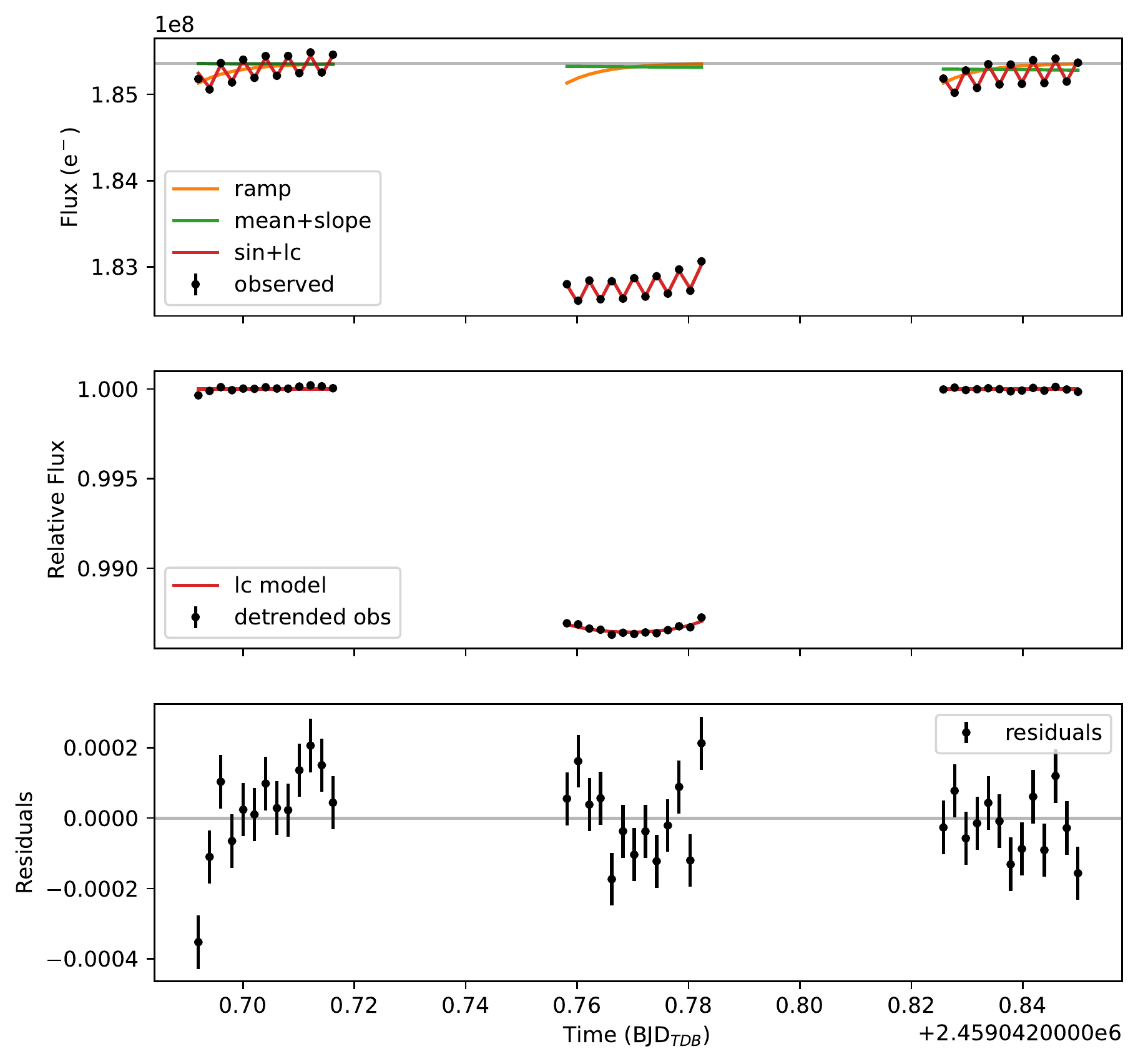}
    \vspace{-1em}
    \caption{The broadband data for the second transit of TOI-674 b. \textit{Top:} The raw transit data, with the systematics and transit model. \textit{Middle:} The detrended transit data and lightcurve model. \textit{Bottom:} The white-lightcurve residuals.}
    %\vspace{-1em}
    \label{fig:t2_white}
\end{figure*}
\begin{figure*}[htb]
    \centering
    \includegraphics[width=0.6\textwidth]{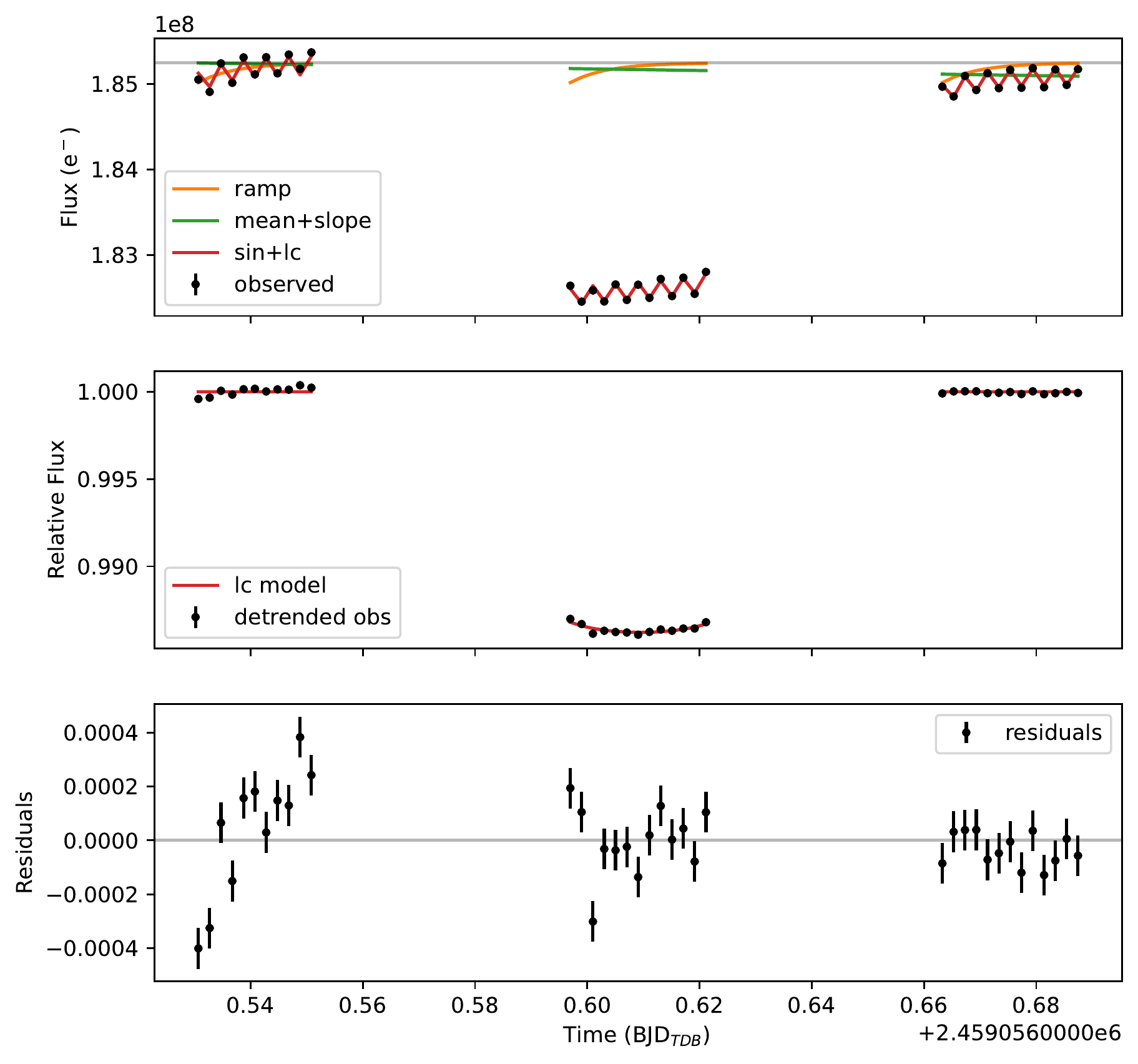}
    \vspace{-1em}
    \caption{The broadband data for the third transit of TOI-674~b. \textit{Top:} The raw transit data, with the systematics and transit model. \textit{Middle:} The detrended transit data and lightcurve model. \textit{Bottom:} The white-lightcurve residuals.}
    %\vspace{-1em}
    \label{fig:t3_white}
\end{figure*}

\begin{figure}[htb]
    \centering
    \includegraphics[width=0.45\textwidth]{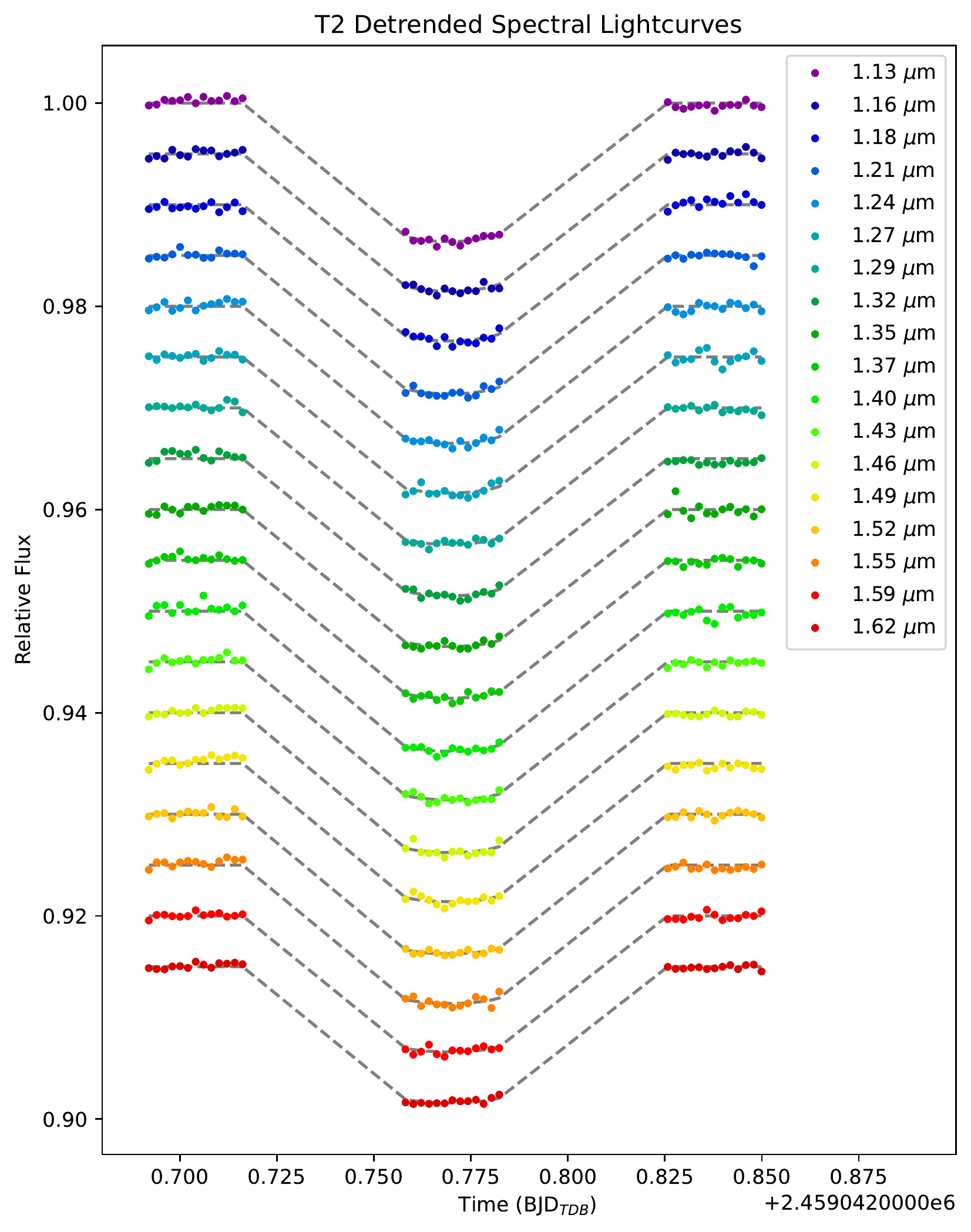}
    \vspace{-1em}
    \caption{Detrended spectral lightcurves and the transit models for the second transit of TOI-674 b.}
    %\vspace{-1em}
    \label{fig:t2_detrend}
\end{figure}
\begin{figure}[htb]
    \centering
    \includegraphics[width=0.45\textwidth]{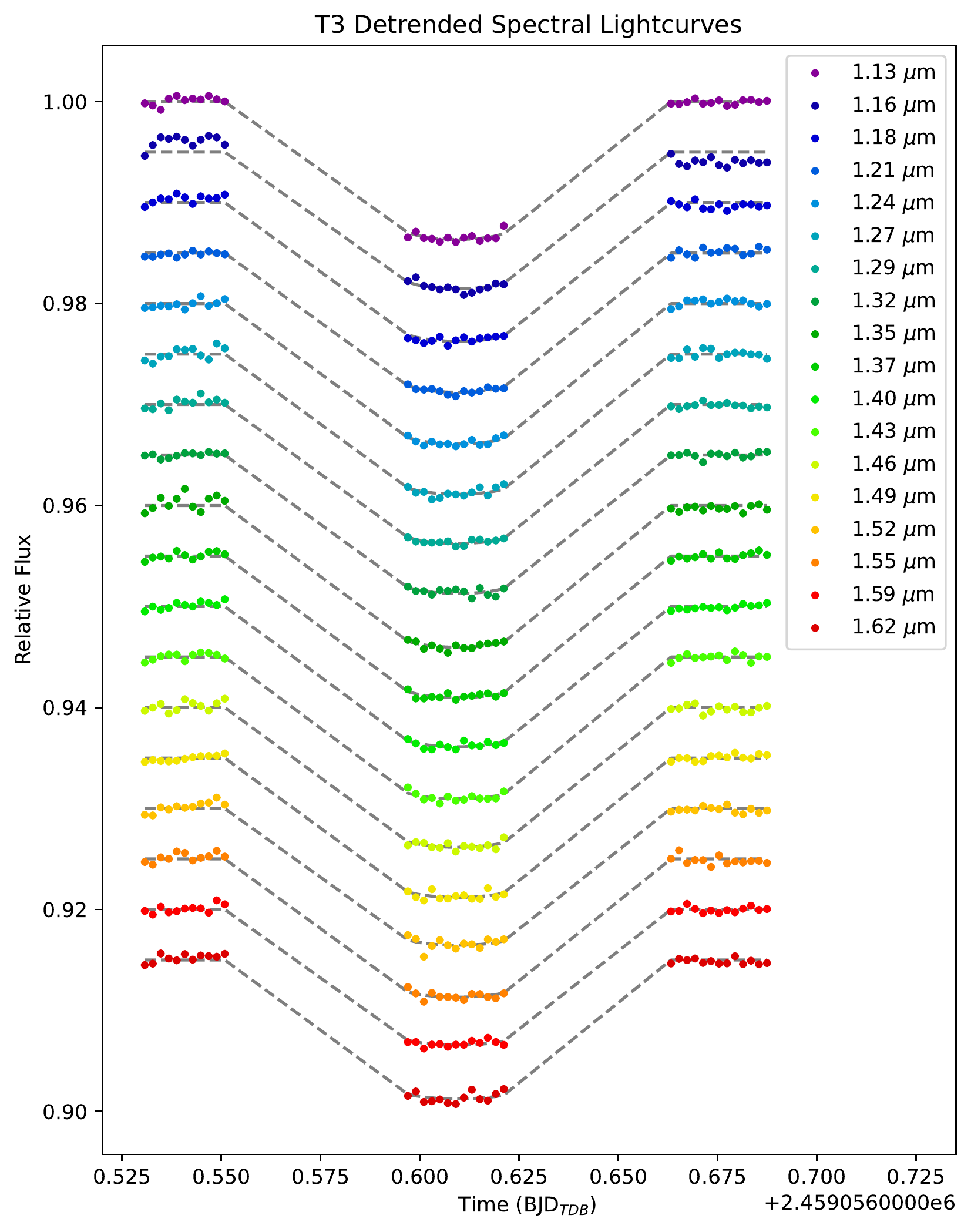}
    \vspace{-1em}
    \caption{Detrended spectral lightcurves and the transit models for the third transit of TOI-674 b.}
    %\vspace{-1em}
    \label{fig:t3_detrend}
\end{figure}

%%%%%%%%%%%%%%%%%%%%%%%%%
%   FULL CORNER GOES HERE
%%%%%%%%%%%%%%%%%%%%%%%%%


\vspace{5mm}
\facilities{HST, TESS, Spitzer, MAST, ExoFOP, Exoplanet Archive}


\software{astropy \citep{astropy2013,astropy2018}, ExoCTK \citep{bourque2021}, \texttt{exoplanet} \citep{dfm2021}, \texttt{starry} \citep[][]{luger2019}, petitRADTRANS \citep{molliere2019, molliere2020}, Iraclis \citep{tsiaras2016a, tsiaras2016b, tsiaras2018}}



%\appendix
%\section{Best-Fit Atmospheric Model Results \label{app:a}}


\bibliography{bibliography}{}
\bibliographystyle{aasjournal}



\end{document}